\documentclass[12pt]{article}
\usepackage{amsmath,cite,epsfig,a4,times,euscript}
\usepackage[text-hat-accent]{euler}
\usepackage{graphics}
\usepackage[usenames]{color}
\renewcommand{\baselinestretch}{1.1}
\newcommand{\myTitle}[1]{\begin{center}{\bf\Huge #1}\\[5ex]\end{center}}
\newcommand{\myAuthor}[1]{\begin{center}{\Large #1}\\[2ex]\end{center}}
\newcommand{\myAffiliation}[1]{\\[1ex]{\it\large #1}}
\newcommand{\myEmail}[1]{}
\newcommand{\myDate}{\begin{center}{\large\today}\\[5ex]\end{center}}
\newcommand{\myAbstract}[1]{\begin{center}\renewcommand{\baselinestretch}{1}{\bf Abstract}\\[2ex]\parbox{0.8\linewidth}{\small\hspace{15pt} #1}\end{center}\vspace{\baselineskip}}
\newcommand{\myReport}[1]{\hspace{\fill} #1}
\newcommand{\myPreprint}[1]{}
\newcommand{\myKeywords}[1]{}

\newcommand{\myScript}[1]{\EuScript{#1}}
\newcommand{\myParagraph}[1]{\paragraph{#1}\hspace{-2ex}}

\newcommand{\fudgeb}{\\[-0.7ex]}

\newcommand{\lid}[2]{#1\!\cdot\!#2}
\newcommand{\slashp}{p\hspace{-6.5pt}/}

\newcommand{\slashk}{k\hspace{-6.5pt}/}
\newcommand{\slashK}{K\hspace{-7.7pt}/}
\newcommand{\slashe}{e\hspace{-6.0pt}/}
\newcommand{\slashr}{r\hspace{-6.0pt}/}
\newcommand{\Appendix}[1]{Appendix~\ref{#1}}   
\newcommand{\Section}[1]{Section~\ref{#1}}

\newcommand{\Equation}[1]{Eq.~(\ref{#1})}

\newcommand{\ie}{{\it i.e.}}

\newcommand{\Amp}{\myScript{A}}
\newcommand{\srac}[2]{{\textstyle\frac{#1}{#2}}}

\newcommand{\imag}{\mathrm{i}}

\newcommand{\vep}{\varepsilon}
\newcommand{\gQCD}{g_\mathrm{S}}
\renewcommand{\AA}[1]{\langle#1\rangle}
\newcommand{\AS}[1]{\langle#1]}
\newcommand{\SA}[1]{[#1\rangle}
\newcommand{\SSS}[1]{[#1]}
\newcommand{\AL}[1]{\langle#1|}
\newcommand{\AR}[1]{|#1\rangle}
\newcommand{\SL}[1]{[#1|}
\newcommand{\SR}[1]{|#1]}
\newcommand{\kapp}{\kappa}
\newcommand{\kapphat}{\hat{\kappa}}
\newcommand{\kstr}{\kappa^*}
\newcommand{\kstrhat}{\hat{\kappa}^*}
\newcommand{\lng}{direction}
\newcommand{\lngs}{directions}
\newcommand{\eikquark}{eikonal quark}
\newcommand{\graph}[3]{\raisebox{-#3ex}{\epsfig{file=graphs/#1.pdf,width=#2ex}}}

\begin{document}

\myReport{IFJPAN-IV-2014-8}
\myPreprint{}\\[2ex]

\myTitle{%
BCFW recursion for off-shell gluons
}

\myAuthor{%
A.~van~Hameren%
\myAffiliation{%
The H.\ Niewodnicza\'nski Institute of Nuclear Physics\fudgeb
Polisch Academy of Sciences\\
Radzikowskiego 152, 31-342 Cracow, Poland%
\myEmail{hameren@ifj.edu.pl}
}
}

\myDate

\myAbstract{%
It is shown how tree-level multi-gluon helicity amplitudes with an arbitrary number of off-shell external gluons can be calculated via BCFW recursion.
Compact expressions for helicity amplitudes for scattering processes of three and four gluons, with up to three of them off-shell, are presented.
Also, maximum-helicity-violating configurations are identified for up to two off-shell gluons, and the expressions for their helicity amplitudes for an arbitrary number of on-shell gluons are derived.
}

\myKeywords{QCD}

%

\section{Introduction\label{Sec:intro}}
Scattering amplitudes are essential ingredients in the calculation of predictions of observables for high-energy collider experiments like at the LHC.
Ideally, they encode the process dependent content entering a calculation, and their contribution is factorized in some form from universal functions describing the colliding particles.
This mechanism is made explicit in so-called factorization formulas and, again ideally, given solid ground with factorization theorems.

One of such factorization approaches goes under the name of {\em high-energy factorization\/} or {\em $k_T$-factorization}~\cite{Gribov:1984tu,Catani:1990eg,Collins:1991ty,Catani:1994sq}, and from the perspective of the scattering amplitudes it is particular because it demands off-shell initial-state partons.
The main issue in this respect is that, in order for scattering amplitudes to be well-defined, they must be gauge invariant, and that it is not {\it a priori\/} clear how to achieve this with only the Lagrangian of Quantum Chromo Dynamics and the Lehmann Symanzik Zimmermann reduction formula at hand.
The latter approach requires all external partons to be on-shell, while $k_T$-factorization involves off-shell external partons.

A manifestly gauge invariant and constructive definition of scattering amplitudes with an arbitrary number of off-shell external gluons has been presented recently in~\cite{Kotko:2014aba}.
The amplitudes are calculated by considering matrix elements of Fourier transforms of straight infinite Wilson line operators associated with the off-shell external gluons.
The starting point of that paper is the observation that in existing approaches to define and construct scattering amplitudes with off-shell gluons~\cite{Lipatov:1995pn,Antonov:2004hh,vanHameren:2012uj,vanHameren:2012if} Wilson lines always appear in one form or another.
In~\cite{vanHameren:2012if} for example, the Wilson lines are represented by eikonal quark lines.
In fact, for explicit calculations, the approach of~\cite{Kotko:2014aba} can be formulated such that each external off-shell gluon should be replaced by a unique quark-anti-quark pair satisfying eikonal Feynman rules.

In~\cite{vanHameren:2012uj}, some explicit compact expressions were given for four-gluon helicity amplitudes with one of them off-shell, and it was observed that they follow the simple structure of maximum-helicity-violating (MHV) amplitudes, well known for on-shell multi-parton amplitudes~\cite{Parke:1986gb,Berends:1987me,Mangano:1990by}.
In~\cite{vanHameren:2013csa} a similar phenomenon was observed for four-parton amplitudes with an off-shell initial-state quark.
Obtaining compact expressions for on-shell multi-parton amplitudes, be it for MHV configurations or non-MHV configurations, was dramatically simplified with the introduction of BCFW recursion~\cite{Britto:2004ap,Britto:2005fq}.
Thus the question naturally arises whether this recursion can be generalized to amplitudes with off-shell partons.
In this write-up, we show that this is indeed possible, despite the seeming contradiction that this recursion is also known as {\em on-shell recursion\/}.
The latter refers to the fact that the recursion happens on the level of gauge-invariant amplitudes, rather than on the level of non-gauge-invariant off-shell currents like Berends-Giele recursion~\cite{Berends:1987me}.
We will see that also the gauge-invariant amplitudes with off-shell partons can be calculated with recursion at the level of only such amplitudes.

The structure of the paper is as follows.
In \Section{Sec:definitions} we give a definition of the amplitudes under consideration in the terms suitable for \Section{Sec:construction}, in which the application of BCFW recursion to amplitudes with off-shell gluons is given.
The connection of the amplitudes with fully on-shell amplitudes is explained in \Section{Sec:limit}, while some explicit calculations are presented in \Section{Sec:examples}.
It involves helicity amplitudes for three-gluon and four-gluon processes, both with up to three off-shell gluons.
In \Section{Sec:MHV} the MHV amplitudes for an arbitrary number of gluons with at most two off-shell are derived.
\Section{Sec:summary}, finally, is the summary closing the paper.

\section{\label{Sec:definitions}Definitions}
A scattering amplitude for a number of $n$ external gluons is a function of $2n$ four-vectors, namely the momenta $k_1,k_2,\ldots,k_n$ and the \lngs\ $p_1,p_2,\ldots,p_n$, satisfying the conditions
%
\begin{align}
k_1^\mu + k_2^\mu + \cdots + k_n^\mu = 0
&\qquad\textrm{momentum conservation}\label{Eq:momcons}\\
p_1^2 = p_2^2 = \cdots = p_n^2 = 0
&\qquad\textrm{light-likeness}\label{momcon1}\\
\lid{p_1}{k_1} = \lid{p_2}{k_2}=\cdots=\lid{p_n}{k_n}=0
&\qquad\textrm{eikonal condition}\label{Eq:eikcon}
\end{align}
%
If a momentum $k^\mu$ is time-like, then the \lng\ $p^\mu$ cannot be real, but it can always be constructed given $k^\mu$ (\Appendix{App:direction}).
For a light-like momentum, the \lng\ is equal to the momentum.
In applications within $k_T$-factorization, the momentum of an initial-state off-shell gluon is space-like, and it is usually defined in terms of the \lng\ and a transverse momentum as $k^\mu=xp^\mu+k_T^\mu$ with $\lid{p}{k_T}=0$ and $x\in[0,1]$.
It turns out that for the amplitude, there is a freedom in the choice of $k_T^\mu$ which is essential in the approach presented here, and that the momentum $k^\mu$ must be considered as the fundamental quantity rather than $k_T^\mu$.
With the help of an auxiliary four-vector $q^\mu$ with $q^2=0$, we may define the transverse vector of the momentum $k^\mu$ by
%
\begin{equation}
k_T^\mu(q) = k^\mu - x(q)p^\mu
\quad\textrm{with}\quad
x(q)\equiv\frac{\lid{q}{k}}{\lid{q}{p}}
~,
\end{equation}
%
so that $k_T^\mu$ satisfies both the relations $\lid{p}{k_T}=0$ and $\lid{q}{k_T}=0$.
Although its components depend on $q^\mu$, its square $k_T^2=k^2$ does not.
The condition that $k_T^\mu$ is transverse to both $p^\mu$ and $q^\mu$ is enough to construct it explicitly with the help of the two ``polarization vectors'' that can naturally be constructed with $p^\mu$ and $q^\mu$.
We may write
%
\begin{equation}
k_T^\mu(q) = -\frac{\kapp}{2}\,\frac{\AS{p|\gamma^\mu|q}}{\SSS{pq}}
             -\frac{\kstr}{2}\,\frac{\AS{q|\gamma^\mu|p}}{\AA{qp}}
\quad\textrm{with}\quad
\kapp = \frac{\AS{q|\slashk|p}}{\AA{qp}}
\;\;,\;\;
\kstr = \frac{\AS{p|\slashk|q}}{\SSS{pq}}
~.
\label{Eq:defkappas}
\end{equation}
%
The spinors are defined in \Appendix{App:spinors}.
If all four-vectors involved are real, then $\kstr$ is the complex conjugate of $\kapp$.
In general, it is not.
What does hold in general, however, is
%
\begin{equation}
k^2 =-\kapp\kstr
~.
\end{equation}
%
It is furthermore important to note that not only their product is independent of $q^\mu$, but
\begin{equation}
\textrm{\em both $\kapp$ and $\kstr$ are independent of the auxiliary momentum $q^\mu$.}
\end{equation}
%
This is shown in \Appendix{App:Schouten} with the help of the Schouten identity.

The off-shell external gluons are represented by Wilson lines in~\cite{Kotko:2014aba}.
In~\cite{vanHameren:2012if} they are represented by auxiliary eikonal quarks.
For the eventual calculation of tree-level amplitudes, these representations are completely equivalent.
For the discussion here, we need to note that the external off-shell gluons are thus represented by two external lines.
Both in~\cite{Kotko:2014aba} and in~\cite{vanHameren:2012if} it is suggested to choose the momentum flow such that the full momentum $k^\mu$ of the off-shell gluon flows in on one end, and momentum identical zero flows in on the other side.
One may, however, add any momentum $q^\mu$ on one side and subtract it on the other side, as long as it satisfies $\lid{p}{q}=0$.
The amplitude will be independent of $q^\mu$.
So instead of \Equation{Eq:eikcon}, we may require that the two momenta $k_{\mathrm{bgn}}^\mu$ and $k_{\mathrm{end}}^\mu$ flowing into the two ends of the Wilson/quark-line satisfy
%
\begin{equation}
\lid{p}{k_{\mathrm{bgn}}}=\lid{p}{k_{\mathrm{end}}}=0
\quad\textrm{and}\quad
k_{\mathrm{bgn}}^\mu+k_{\mathrm{end}}^\mu=k^\mu
\label{Eq:bgnend}
\end{equation}
%
for each external off-shell gluon.

In this write-up, we will consider the calculation of {\em color-ordered\/} or {\em dual\/} amplitudes, like in~\cite{Kotko:2014aba}.
These contain only planar Feynman graphs, and are composed with color-stripped Feynman rules.
In the graphical representation, external gluons will be represented with double lines here.
If we wish to specify that they are on-shell, we represent them with thick solid lines.
Internal, virtual, gluons will also be represented by thick lines.
The double lines can be bent apart, to form a single \eikquark\ line.
As an example, the planar Feynman graphs contributing to the color-ordered amplitude for the process $\emptyset\to g^*g^*g$ are 
%
\begin{equation}
\graph{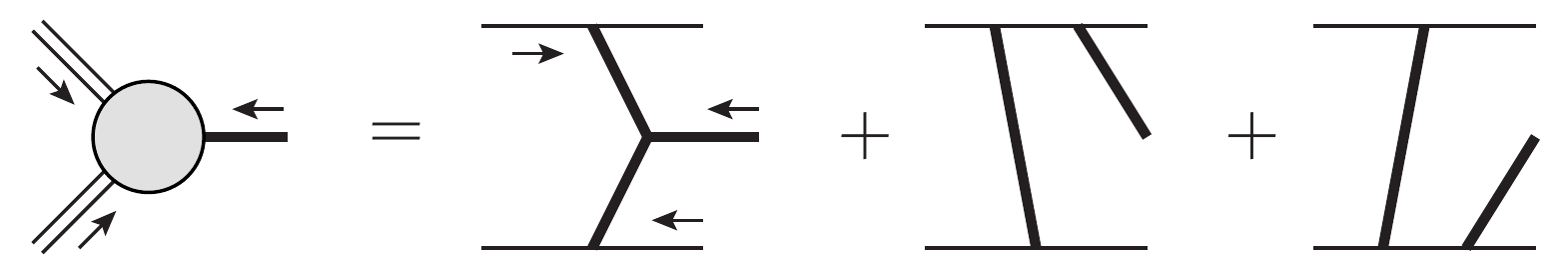}{50}{3.8}
\label{Eq:Lipvtx}
\end{equation}
%
The arrows indicate the flow of momentum.
Notice that they stay at the same side of the same \eikquark\ line once the double line indicating the external off-shell gluon is bent open.
%
%
The Feynman rules, in the Feynman gauge, are as follows:
%
\begin{align}
\graph{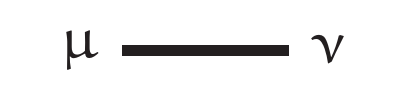}{13}{0.7} &= \frac{-\eta^{\mu\nu}}{K^2}
\quad\quad\quad
\graph{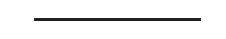}{7}{0} = \frac{1}{2\lid{p}{K}}
\quad\quad\quad
\graph{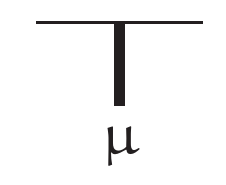}{7}{2.5} = \sqrt{2}\,p^\mu
\notag\\
\graph{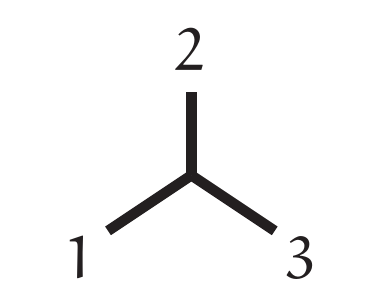}{12}{3.5} &= \frac{1}{\sqrt{2}}
                         \,\big[(K_1-K_2)^{\mu_3}\eta^{\mu_1\mu_2}
                               +(K_2-K_3)^{\mu_1}\eta^{\mu_2\mu_3}
                               +(K_3-K_1)^{\mu_2}\eta^{\mu_3\mu_1}\big]
\\
\graph{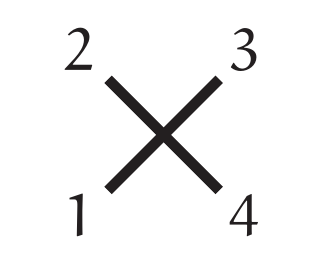}{10.5}{3.5} &= \frac{-1}{2}\,\big[
                            2\,\eta^{\mu_1\mu_3}\eta^{\mu_2\mu_4}
                            -\eta^{\mu_1\mu_2}\eta^{\mu_3\mu_4}
                            -\eta^{\mu_1\mu_4}\eta^{\mu_2\mu_3}\big]
\notag
\end{align}
%
In these rules, the symbol $p$ refers to the \lng\ associated with the \eikquark\ line, the symbol $K$ refers to momentum flowing through a propagator or into a vertex, and $\eta^{\mu\nu}$ is the metric tensor.
We set the coupling constant $\gQCD$ to unity for convenience.
We removed the imaginary unit from the numerator compared to rules in~\cite{Kotko:2014aba}.
Also the factors of $\sqrt{2}$ in the eikonal vertex and propagator are different, and more suitable for our purpose.
Amplitudes in the convention of~\cite{Kotko:2014aba} are obtained by multiplying our amplitudes with a factor
%
\begin{equation}
\frac{\imag\,(-1)^{1+(\textrm{\# on-shell gluons})}}
     {2^{(\textrm{\# off-shell gluons})}}
~.
\end{equation}
%
Applying the rules to $\emptyset\to g^*g^*g$, we find
%
\begin{align}
\graph{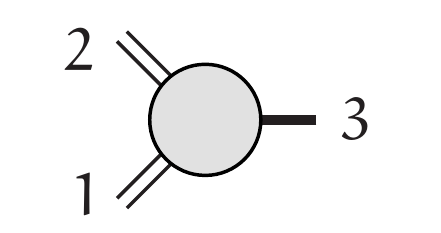}{14.5}{3.2}&=
\sqrt{2}\,
\frac{ \lid{(k_1-k_2)}{\vep_3}\,\lid{p_1}{p_2}
      -\lid{(k_2-k_3)}{p_1}\,\lid{p_2}{\vep_3}
      -\lid{(k_3-k_1)}{p_2}\,\lid{p_1}{\vep_3}}
     {k_1^2\,k_2^2}
\notag\\
&-\sqrt{2}\,
  \frac{\lid{p_1}{p_2}\,\lid{p_1}{\vep_3}}{k_2^2\,(-\lid{p_1}{k_3})}
-\sqrt{2}\,
\frac{\lid{p_1}{p_2}\,\lid{p_2}{\vep_3}}{k_1^2\,\lid{p_2}{k_3}}
~.
\end{align}
%
Inserting $\vep_3\leftarrow k_3$ and using momentum conservation, so that $2\lid{k_1}{k_3}=k_2^2-k_1^2$, we find that the expression vanishes, as it should according to the Ward identity, confirming the consistency of the Feynman rules.

It needs to be stressed that the Feynman rules above are not applied explicitly in the calculation of amplitudes in the following.
The proof of BCFW recursion is based on the observation that the amplitudes can be represented as sums of Feynman graphs following the Feynman rules.
In the recursion itself, they do not appear.

\section{\label{Sec:construction}Construction}
The BCFW approach~\cite{Britto:2004ap,Britto:2005fq} is based on the simple observation that for a rational function $f$ of a complex variable $z$ which vanishes at infinity, we have
\begin{equation}
\lim_{z\to\infty}f(z)=0
\quad\Rightarrow\quad
\oint\frac{dz}{2\pi\imag}\,\frac{f(z)}{z}=0
~,
\end{equation}
%
where the integration contour expands to infinity and necessarily encloses all poles of $f$.
This directly leads to the relation
%
\begin{equation}
f(0) = \sum_i\frac{\lim_{z\to z_i}f(z)(z-z_i)}{-z_i}
\label{Eq:contour}
~,
\end{equation}
%
where the sum is over all poles of $f$, and $z_i$ is the position of pole number $i$.
%

\subsection{Shifted momenta}
To apply the observation, the scattering amplitude needs to be turned into a function of a complex variable $z$, whose value at $z=0$ is the desired quantity.
This can be achieved by deforming the external momenta into complex space while still satisfying the relations (\ref{Eq:momcons}) and (\ref{Eq:eikcon}).
This can be done as follows.
We choose two external gluons $i$ and $j$, and use the \lng\ of one as the auxiliary momentum to define the transverse momentum of the other, and vice versa, so
%
\begin{align}
k_i^\mu &= x_i(p_j)p_i^\mu
         - \frac{\kapp_i}{2}\,\frac{\AS{i|\gamma^\mu|j}}{\SSS{ij}}
         - \frac{\kstr_i}{2}\,\frac{\AS{j|\gamma^\mu|i}}{\AA{ji}}
\\
k_j^\mu &= x_j(p_i)p_j^\mu
         - \frac{\kapp_j}{2}\,\frac{\AS{j|\gamma^\mu|i}}{\SSS{ji}}
         - \frac{\kstr_j}{2}\,\frac{\AS{i|\gamma^\mu|j}}{\AA{ij}}
~.
\end{align}
%
We adopt the notation $\AR{i}\equiv\AR{p_i}$, $\AR{j}\equiv\AR{p_j}$ etc.\ here.
Next, we choose the so-called {\em shift vector\/}
%
\begin{equation}
e^\mu\equiv\srac{1}{2}\AS{i|\gamma^\mu|j}
~,
\end{equation}
%
which has the essential properties that
%
\begin{equation}
\lid{p_i}{e} = \lid{p_j}{e} = \lid{e}{e} = 0
~.
\end{equation}
%
Using this vector, we then define the shifted momenta
%
\begin{align}
\hat{k}_i^\mu(z) &\equiv k_i^\mu+ze^\mu
        = x_i(p_j)p_i^\mu
         - \frac{\kapp_i-\SSS{ij}z}{2}\,\frac{\AS{i|\gamma^\mu|j}}{\SSS{ij}}
         - \frac{\kstr_i}{2}\,\frac{\AS{j|\gamma^\mu|i}}{\AA{ji}}
\\
\hat{k}_j^\mu(z) &\equiv k_j^\mu-ze^\mu
        = x_j(p_i)p_j^\mu
         - \frac{\kapp_j}{2}\,\frac{\AS{j|\gamma^\mu|i}}{\SSS{ji}}
         - \frac{\kstr_j+\AA{ij}z}{2}\,\frac{\AS{i|\gamma^\mu|j}}{\AA{ij}}
~.
\end{align}
%
Replacing the momenta $k_i^\mu,k_j^\mu$ with $\hat{k}_i^\mu(z),\hat{k}_j^\mu(z)$ in the amplitude obviously does not violate momentum conservation, since we add the same vector to one momentum as we subtract from the other.
Also, we still have $\lid{p_i}{\hat{k}_i(z)}=0$ and  $\lid{p_j}{\hat{k}_j(z)}=0$, so all necessary conditions are fulfilled.
Notice that the eventual effect of shifting the momenta is that the values of $\kapp_i$ and $\kstr_j$ shift
while $\kstr_i$ and $\kapp_j$ are not affected.

We could also have chosen $e^\mu=\srac{1}{2}\AS{j|\gamma^\mu|i}$, and shift $\kstr_i$ and $\kapp_j$ instead.
The choice of the shift vector only matters if one or both external gluons are on-shell.
Then, the shift vector should be chosen to match the helicity of the on-shell gluon, as prescribed in the original BCFW approach, with the consequence that if both gluons are on-shell, they should have opposite helicity.
For completeness, we repeat the shifts including on-shell gluons
%
\begin{align}
\hat{k}_i^\mu &= k_i^\mu + \srac{z}{2}\AS{i|\gamma^\mu|j}
&
\hat{k}_j^\mu &= k_j^\mu - \srac{z}{2}\AS{i|\gamma^\mu|j}
\notag\\
\textrm{$i$ off-shell:}\quad\kapphat_i &= \kapp_i-z\SSS{ij}
&
\textrm{$j$ off-shell:}\quad\kstrhat_j &= \kstr_j+z\AA{ij}
\label{Eq:shifts}\\
\textrm{$i$ on-shell $-$hel:}\quad\SR{\hat{i}} &= \SR{i} + z\SR{j}
&
\textrm{$j$ on-shell $+$hel:}\quad\AR{\hat{j}} &= \AR{j} - z\AR{i}
\notag
\end{align}
%
It has to be stressed that only in the case of on-shell gluons, when the momentum and the \lng\ are identical, the \lng\ and one of its spinors shift, and that
\begin{equation}
\textrm{\em for off-shell external gluons, the \lngs\ do not shift\/}.
\end{equation}
%
As mentioned before, we will only consider color-ordered amplitudes.
Also, we will only consider the case when gluon $i$ and $j$ are adjacent, and without loss of generality, we will label them with $1$ and $n$, where $n$ is the total number of gluons.
So the function of the complex variable $z$ whose value at $z=0$ gives the desired amplitude is given by
%
\begin{equation}
\hat{\Amp}(z) \equiv \Amp(\hat{k}_1(z),k_{2},\ldots,k_{n-1},\hat{k}_n(z))
~,
\end{equation}
%
where $\hat{\Amp}(0)=\Amp(k_1,k_{2},\ldots,k_{n-1},k_n)$ is the color-ordered $n$ gluon amplitude, with any of them off-shell.
%

\subsection{Behavior at $z\to\infty$}
The first issue that needs to be addressed it the behavior of $\hat{\Amp}(z)$ for $z\to\infty$.
$\hat{\Amp}$ is a rational function of $z$, and denominator factors come from the propagators in the graphs contributing to the amplitude.
Numerator factors can only come from triple gluon vertices.
Following the reasoning in~\cite{Britto:2005fq}, in the worst case there will be one more triple gluon vertex contributing a power of $z$ to the numerator than there will be propagators contributing powers of $z$ to the denominator.
For on-shell gluons, however, each of the polarization vectors of the shifted gluons also contribute a power of $z$ to the denominator, so eventually, in the worst case, $\hat{A}(z)\propto z^{-1}$ for $z\to\infty$.
We see that, in order for this argumentation to hold in case of off-shell gluons, we must include their propagators in the amplitude under consideration, so they contribute powers of $z$ to the denominator via the shift of $\kapp$ or $\kstr$.
This means that here,
%
\begin{equation}
\textrm{\em we consider amplitudes before they are multiplied by a factor proportional to $\sqrt{|k_i^2|}$}
\end{equation}
for each off-shell gluon.
Such a factor was included in the definition of the amplitude in~\cite{vanHameren:2012if} for example, to arrive at the correct on-shell limit if the virtuality of the off-shell gluons is taken to zero.
We will return to this issue in \Section{Sec:limit}.

The graphs contributing to the amplitude do not only contain gluon propagators and vertices, but also \eikquark\ propagators and vertices.
The vertices are independent of the momenta, and the propagators can only contribute powers of $z$ to the denominator, so these do not spoil the argumentation above.
Notice that the \eikquark\ lines of shifted off-shell gluons are not influenced at all by the shift of the momenta, because the shift vector satisfies $\lid{p_1}{e}=\lid{p_n}{e}=0$.
Only the denominators of other \eikquark\ lines may depend on $z$.
%

\subsection{Contributing poles}
The poles in $\hat{\Amp}(z)$ come from the propagator denominators that depend on $z$.
A graphical representation of all possible poles is given below.
It constitutes \Equation{Eq:contour}.
%
\begin{equation}
\graph{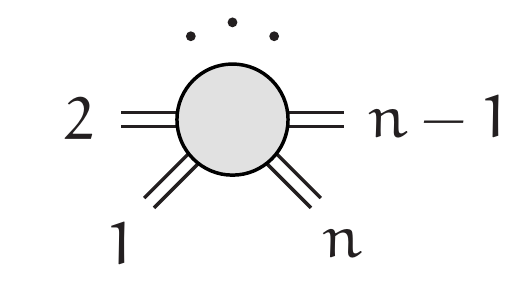}{17}{4.5}
\;=\;\sum_{i=2}^{n-2}\sum_{h=+,-}\mathrm{A}_{i,h}
\;+\;\sum_{i=2}^{n-1}\mathrm{B}_{i}
\;+\;\mathrm{C}\;+\;\mathrm{D}
~,
\end{equation}
%
where
%
\begin{align}
\mathrm{A}_{i,h} &= \graph{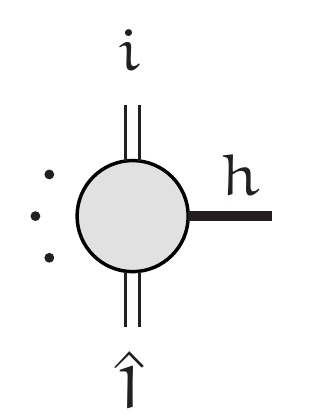}{10.3}{5.8}\frac{1}{K_{1,i}^2}
           \graph{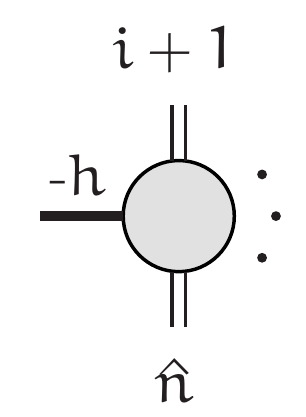}{10}{6}
\quad&
\mathrm{B}_{i} &= \graph{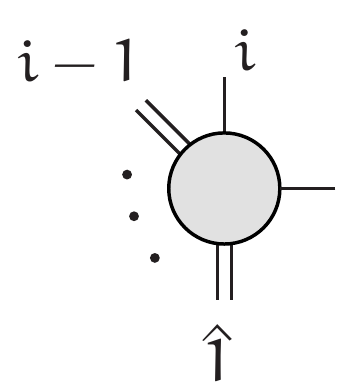}{11.7}{5.4}\frac{1}{2\lid{p_i}{K_{i,n}}}
         \graph{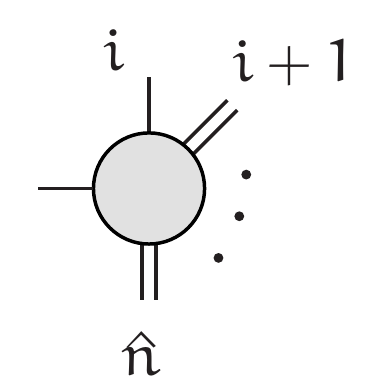}{12}{5.4}
\notag\\
\mathrm{C} &= \frac{1}{\kapp_1}\graph{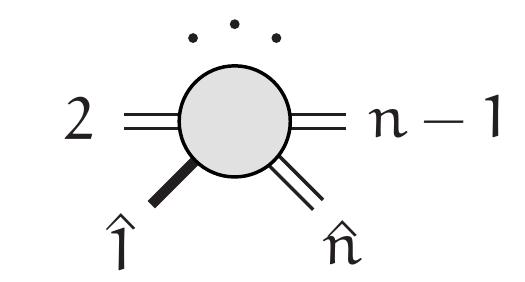}{17}{4.5}
\quad&
\mathrm{D} &= \frac{1}{\kstr_n}\graph{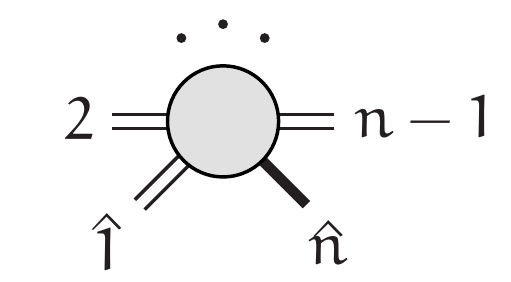}{17}{4.5}
\end{align}
The hatted numbers label the shifted external gluons.
Remember that we use the convention that double lines may refer to both off-shell and on-shell external gluons, and that only when we wish to specify that an external gluon is on-shell, we represent it by a thick solid line.
For internal momenta, we adopt the notation
%
\begin{equation}
K_{i,j}^\mu \equiv k_{i}^\mu + k_{i+1}^\mu+\cdots+k_j^\mu 
~.
\end{equation}
%

\myParagraph{The terms A$_{i,h}$} correspond to the usual contribution that also appears in the BCFW construction for on-shell gluons.
A pole appears when the denominator $\hat{K}_{i,n}^2(z)$ of the propagator of an internal gluon with shifted momentum vanishes, \ie\ when this internal gluon becomes on-shell, which happens for the value
%
\begin{equation}
z = -\frac{K_{i,n}^2}{2\lid{e}{K_{i,n}}}
  = -\frac{K_{i,n}^2}{\AS{1|\slashK_{i,n}|n}}
~.
\end{equation}
%
The two blobs therefor represent well-defined amplitudes with an on-shell gluon referred to by the thick line.
The sum over $h$ is the sum over the possible helicities of these on-shell gluons.
This sum implies a minus sign, which is missing in the inserted propagator.
The explicit denominator $K_{1,i}^2$ corresponds to the $-z_k$ in the denominator of \Equation{Eq:contour}.

\myParagraph{The terms B$_i$} refer to the situation when the denominator of the propagator of an \eikquark\ line vanishes.
This means that $\lid{p_i}{\hat{K}_{i,n}(z)}=0$, where $p_i^\mu$ is the \lng\ associated with the \eikquark\ line and $\hat{K}_{i,n}^\mu(z)$ is the shifted momentum flowing through the propagator.
This happens for the value
%
\begin{equation}
z = -\frac{2\lid{p_i}{K_{i,n}}}{2\lid{p_i}{e}}
  = -\frac{2\lid{p_i}{K_{i,n}}}{\AA{1i}\SSS{in}}
~.
\end{equation}
%
This again means that, according to \Equation{Eq:bgnend}, the two blobs represent well-defined amplitudes both with an off-shell gluon with \lng\ $p_i^\mu$.
This term is absent if $i$ labels an on-shell external gluon.

\myParagraph{The term C} is only present if $1$ labels an off-shell gluon.
It appears due to the pole when the square $\hat{k}_1^2(z)$ of the external momentum vanishes, so that the gluon becomes on-shell.
The square $\hat{k}_1^2(z)$ vanishes because $\hat{\kapp}_1(z)$ vanishes, so we have
%
\begin{equation}
\hat{k}_1^\mu \equiv \hat{k}_1^\mu\big(z=\kapp_1/\SSS{1n}\big)
 = x_1(p_n)p_1^\mu 
              - \frac{\kstr_1}{2}\,\frac{\AS{n|\gamma^\mu|1}}{\AA{n1}}
\label{Eq:kTermC}
~.
\end{equation}
%
All graphs contributing to C have the propagator with momentum $\hat{k}_1^\mu$, which is connected to the only eikonal vertex with $p_1^\mu$ each of these graphs contains.
This means that this vertex can be considered to be contracted with a current satisfying current conservation
%
\begin{equation}
\mathrm{C}
=
\frac{1}{\kapp_1}\graph{graph08}{17}{4.5}
=
\sqrt{2}\,p_1^\mu\,\frac{1}{\kapp_1\kstr_1}\,J_\mu
\qquad\textrm{where}\qquad
\hat{k}_1^\mu\,J_\mu = 0
~.
\end{equation}
%
We took the propagator outside of $J_\mu$.
Using \Equation{Eq:kTermC}, we thus find
%
\begin{equation}
\mathrm{C} = \frac{1}{x_1(p_n)\kapp_1}\,\vep_1^\mu\,J_\mu
\quad,\quad
\vep_1^\mu = \frac{\AS{n|\gamma^\mu|1}}{\sqrt{2}\,\AA{n1}}
~.
\end{equation}
%
So C is given by the amplitude for which gluon $1$ is on-shell and has helicity $+$, the {\em opposite\/} of the helicity suggested by the shift vector $\srac{1}{2}\AS{1|\gamma^\mu|n}$.
Remember that its momentum is $\hat{k}_1^\mu$ and not $p_1^\mu$.
Still we do have $\lid{\hat{k}_1}{\vep_1}=0$.
%
Using the explicit expression $x_1(p_n)=\lid{p_n}{k_1}/\lid{p_n}{p_1}$ and strategic choices for the auxiliary vectors for $\kapp_1,\kstr_1,\kstr_n$, we get the shifted quantities
%
\begin{equation}
\AR{\hat{k}_1} = \frac{\slashk_1\SR{n}}{\sqrt{x_1}\SSS{1n}}
\;\;,\;\;
\SR{\hat{k}_1} = \sqrt{x_1}\,\SR{1}
\quad,\quad
\kstrhat_n = \frac{\AS{n|\slashk_n+\slashk_1|1}}{\SSS{n1}}
\;\;\textrm{or}\;\;
\AR{\hat{n}} = \frac{(\slashp_n+\slashk_1)\SR{1}}{\SSS{n1}}
~,
\label{Eq:Cbraket}
\end{equation}
%
where the latter option depends on whether gluon $n$ is off-shell, or on-shell with helicity $+$.

\myParagraph{The term D} is only present if $n$ labels an off-shell gluon.
Similarly to C, it is given by the amplitude for which gluon $n$ is on-shell and has helicity~$-$,
%
\begin{equation}
\mathrm{D} = \frac{1}{x_n(p_1)\kstr_n}\,\vep_n^\mu\,J_\mu
\quad,\quad
\vep_n^\mu = \frac{\AS{n|\gamma^\mu|1}}{\sqrt{2}\,\SSS{n1}}
~.
\end{equation}
%
Now the shifted quantities are
%
\begin{equation}
\SR{\hat{k}_n} = \frac{\slashk_n\AR{1}}{\sqrt{x_n}\AA{n1}}
\;\;,\;\;
\AR{\hat{k}_n} = \sqrt{x_n}\,\AR{n}
\quad,\quad
\kapphat_1 = \frac{\AS{n|\slashk_1+\slashk_n|1}}{\AA{n1}}
\;\;\textrm{or}\;\;
\SR{\hat{1}} = \frac{(\slashp_1+\slashk_n)\AR{n}}{\AA{1n}}
~,
\label{Eq:Dbraket}
\end{equation}
%
where the latter option depends on whether gluon $n$ is off-shell, or on-shell with helicity $-$.

\myParagraph{The situation for B$_2$ (and B$_{n-1}$)} needs some special attention in case gluon $1$ is on-shell, because it involves a two-point amplitude that is only defined for shifted momenta.
Let us consider the case in which gluon $1$ has helicity~$+$.
We have
%
\begin{align}
\graph{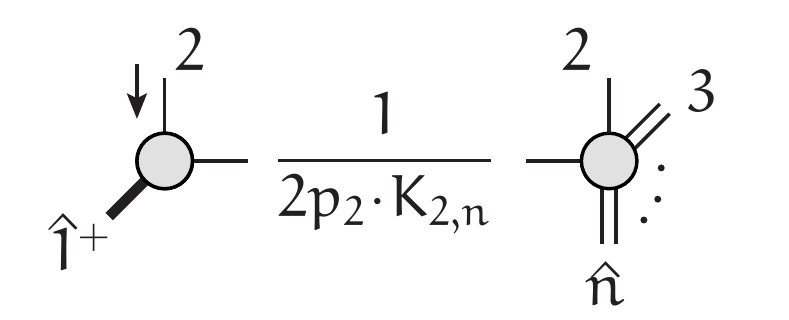}{27}{5}
= \frac{\AS{q|\gamma_\mu|\hat{1}}}{\sqrt{2}\AA{q\hat{1}}}\,\sqrt{2}\,p_2^\mu
  \,\frac{-1}{2\lid{p_2}{k_1}}\graph{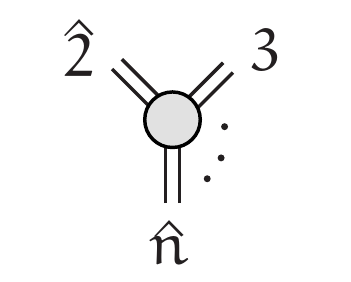}{12}{4.5}
~,
\end{align}
%
off-shell gluon $2$ has momentum $\hat{K}_{1,2}$ which satisfies $\lid{p_2}{\hat{K}_{1,2}}=0$, where $q^\mu$ is an auxiliary momentum used to construct the polarization vector for on-shell gluon $1$, and where we used $\lid{p_2}{K_{2,n}=-\lid{p_2}{(k_1+k_2)}}=-\lid{p_2}{k_1}$.
The expression seems to depend on $q^\mu$, but we will see that the dependence disappears.
The shifted spinors are found as follows.
The shift vector is $\frac{1}{2}\AS{n|\gamma^\mu|1}$, so
%
\begin{equation}
\lid{p_2}{\hat{K}_{1,2}(z)}=0
\quad\Leftrightarrow\quad
\lid{p_2}{k_1} + \frac{z}{2}\AA{n2}\SSS{21}=0
\quad\Leftrightarrow\quad
z = -\frac{\AA{12}}{\AA{n2}}
~,
\end{equation}
%
and
%
\begin{equation}
\SR{\hat{1}}=\SR{1}
\quad,\quad
\AR{\hat{1}} = \AR{1} - \frac{\AA{12}}{\AA{n2}}\,\AR{n}
             = \frac{\AA{1n}}{\AA{2n}}\,\AR{2}
~,
\end{equation}
%
where we used the Schouten identity in the last step.
Substiting this, we find
%
\begin{equation}
\frac{\AS{q|\gamma_\mu|\hat{1}}}{\sqrt{2}\AA{q\hat{1}}}\,\sqrt{2}\,p_2^\mu
  \,\frac{-1}{2\lid{p_2}{k_1}}
=\frac{-\AA{q2}\SSS{2\hat{1}}}{\AA{q\hat{1}}\AA{21}\SSS{12}}
=\frac{-\AA{2n}\AA{q2}\SSS{21}}{\AA{1n}\AA{q2}\AA{21}\SSS{12}}
= \frac{\AA{2n}}{\AA{21}\AA{1n}}
~.
\end{equation}
%
The momentum $\hat{K}_{1,2}^\mu$ is given by
%
\begin{equation}
\hat{K}_{1,2}^\mu = \srac{1}{2}\AS{\hat{1}|\gamma^\mu|\hat{1}} + k_2^\mu
= k_2^\mu + \frac{\AA{1n}}{\AA{2n}}\,\frac{\AS{2|\gamma^\mu|1}}{2}
~.
\end{equation}
%
Using $k_1^\mu=p_1^\mu$ as the auxiliary vector to define the transverse components of $k_2^\mu$, we get
%
\begin{equation}
\hat{K}_{1,2}^\mu 
= x_2(p_1)p_2^\mu - \frac{\kapp_2-\frac{\AA{1n}}{\AA{2n}}\SSS{21}}{2}
                    \,\frac{\AS{2|\gamma^\mu|1}}{\SSS{21}}
- \frac{\kstr_2}{2}\,\frac{\AS{1|\gamma^\mu|2}}{\AA{12}}
~,
\end{equation}
%
and we see that $\hat{K}_{1,2}^\mu=\hat{k}_2^\mu$ defined through the shift of $\kapp_2$.
%
%
For gluon $n$, the shift can also easily be calculated.
So we may summarize
%
\begin{gather}
\graph{graph17}{27}{5}
=  \frac{\AA{2n}}{\AA{21}\AA{1n}}\graph{graph23}{12}{4.5}
\label{Eq:Btwo}\\
\kapphat_2 = \frac{\AS{n|\slashk_2+\slashp_1|2}}{\AA{n2}}
\quad,\quad
\kapphat_n = \frac{\AS{2|\slashk_n+\slashp_1|n}}{\AA{2n}}
\;\;\textrm{or}\;\;
\SR{\hat{n}} = \frac{(\slashp_n+\slashp_1)\AR{2}}{\AA{n2}}
\label{Eq:Btwo2}
~,
\end{gather}
%
where the last option depends on whether gluon $n$ is off-shell, or on-shell with helicity $-$.
For the other helicity for gluon $1$, we find
%
\begin{gather}
\graph{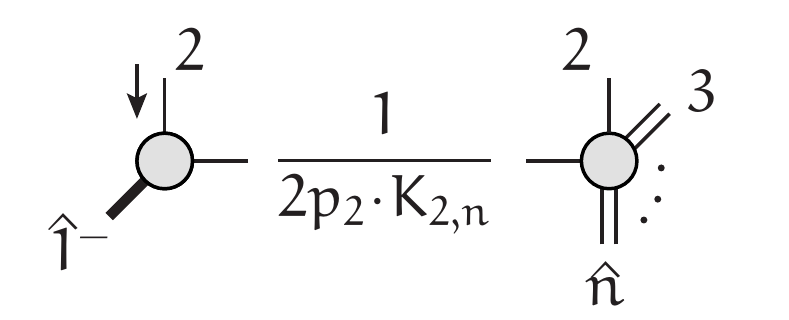}{27}{5}
= \frac{\SSS{n2}}{\SSS{21}\SSS{1n}}\graph{graph23}{12}{5}
\\
\kstrhat_2=\frac{\AS{2|\slashk_2+\slashp_1|n}}{\SSS{2n}}
\quad,\quad
\kstrhat_n = \frac{\AS{n|\slashk_n+\slashp_1|2}}{\SSS{n2}}
\;\;\textrm{or}\;\;
\AR{\hat{n}} = \frac{(\slashk_n+\slashp_1)\SR{2}}{\SSS{n2}}
~,
\end{gather}
%
where the last option depends on whether gluon $1$ is off-shell, or on-shell with helicity $+$.

It is worthwhile to include a general formula for the case that gluon $1$ is off-shell.
%
\begin{equation}
\graph{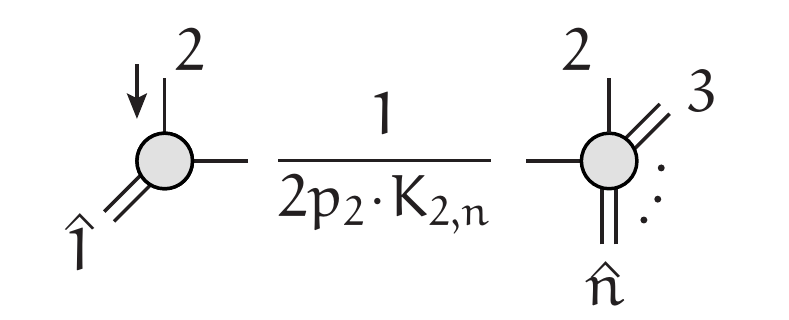}{27}{5}
= \frac{2\lid{p_1}{p_2}}{\kapphat_1\kstr_1}\,\frac{-1}{2\lid{p_2}{k_1}}
\graph{graph23}{12}{4.5}
~.
\end{equation}
%
For shift vector $\frac{1}{2}\AS{1|\gamma^\mu|n}$, we have
%
\begin{equation}
\kapphat_1 = \kapp_1-\SSS{1n}z
\quad,\quad
z = -\frac{2\lid{p_2}{k_1}}{\AA{12}\SSS{2n}}
~,
\end{equation}
%
leading to
%
\begin{equation}
\graph{graph24}{27}{5}
= \frac{\AA{12}^2\SSS{2n}}{\kstr_1\AS{2|\slashk_1|n}}\,\frac{1}{2\lid{p_2}{k_1}}
\graph{graph23}{12}{4.5}
\label{Eq:B2off1}
~.
\end{equation}
%
Now, both $\kapp_2$ and $\kstr_2$ shift:
%
\begin{equation}
\kstrhat_2 = \frac{\AS{2|\slashk_2+\slashk_1|n}}{\SSS{2n}}
\;\;,\;\;
\kapphat_2 = \frac{\AS{1|\slashk_2+\slashk_1|2}}{\AA{12}}
\quad,\quad
\kstrhat_n = \kstr_n+z\AA{1n}
\;\;\textrm{or}\;\;
\AR{\hat{n}} = \AR{n} - z\AR{1}
\end{equation}
%
with $z$ as given before, and with $\kstrhat_n$ or $\AR{\hat{n}}$ depending on whether gluon $n$ is off-shell or on-shell.
with shift vector $\frac{1}{2}\AS{n|\gamma^\mu|1}$, we get
%
\begin{equation}
\graph{graph24}{27}{5}
= \frac{\SSS{21}^2\AA{n2}}{\kapp_1\AS{n|\slashk_1|2}}\,\frac{1}{2\lid{p_2}{k_1}}
\graph{graph23}{12}{4.5}
\label{Eq:B2off2}
~.
\end{equation}
%
with $z=-2\lid{p_2}{k_1}/\AA{n2}/\SSS{21}$ and
%
\begin{equation}
\kapphat_2 = \frac{\AS{n|\slashk_2+\slashk_1|2}}{\AA{n2}}
\;\;,\;\;
\kstrhat_2 = \frac{\AS{2|\slashk_2+\slashk_1|1}}{\SSS{21}}
\quad,\quad
\kapphat_n = \kapp_n+z\SSS{n1}
\;\;\textrm{or}\;\;
\SR{\hat{n}} = \SR{n} - z\SR{1}
~.
\end{equation}
%

\section{\label{Sec:limit}On-shell limits}
The amplitudes for off-shell gluons were defined in~\cite{vanHameren:2012if} such that they match the amplitudes for on-shell gluons in the limit in which the virtuality of the off-shell gluons vanishes.
They are matched such that the square of the off-shell amplitude integrated over the angle, remaining after the magnitude of the virtuallity is taken to zero, is equal to the square of the on-shell amplitude summed over the helicities of the on-shell gluon.
The same relation holds here too.

The expression for an amplitude obtained with the approach suggested in this paper consists of a number of terms which, for each off-shell gluon $j$, can be divided into the following three groups: terms that have an explicit single factor $\kapp_j$ in the denominator, terms that have a single factor $\kstr_j$ in the denominator, and terms that have neither:
\begin{equation}
\Amp(k_j) = \frac{1}{\kstr_j}\,U(k_j) + \frac{1}{\kapp_j}\,V(k_j) + W(k_j)
\label{Eq:limit1}
~.
\end{equation}
%
In the limit in which the virtuality of the off-shell gluon vanishes, both $\kapp_j$ and $\kstr_j$ vanish, and as mentioned earlier, the amplitude has to be multiplied by a factor proportional to the square root of the virtuality before taking the limit.
We may choose this factor to be $\kstr_j$.
So then, the amplitude can be written as
\begin{equation}
\kstr_j\Amp(k_j) = U(k_j) + \frac{\kstr_j}{\kapp_j}\,V(k_j) + \kstr_j\,W(k_j)
~.
\end{equation}
%
This also holds after the amplitude is dressed up with color.
The third term obviously vanishes in the on-shell limit, while the ratio $\kstr_j/\kapp_j$ is independent of the magnitude of the virtuality, and can be parametrized by and angle.
Setting $x_j=\lid{q_j}{k_j}/\lid{q_j}{p_j}=1$ for convenience, we have
%
\begin{equation}
\big|\kstr_j\Amp(k_j)\big|^2 \;\overset{k_j^2\to0}{\longrightarrow}\;
\big|U(p_j)\big|^2 + \big|V(p_j)\big|^2 + e^{2\imag\varphi_j}U(p_j)V(p_j)^* + e^{-2\imag\varphi_j}U(p_j)^*V(p_j)
~.
\end{equation}
%
We see that the interference terms vanish upon integration over the angle $\varphi_j$, and that the result consists of two squares.

In order to identify the helicity of each of the terms, we take the limit in two steps.
First we shift the momentum $k_j^\mu$ following \Equation{Eq:kTermC}, so that $\kapphat_j(z)=0$ and it becomes light-like.
For the second momentum that needs to be shifted we choose any other off-shell gluon, or any on-shell gluon with helicity $+$.
Following the argumentation below \Equation{Eq:kTermC}, we see that gluon $j$, which is now on-shell, will have helicity $+$, while the amplitude will carry an explicit factor $\kapp_j$ in the denominator.
This helicity does not change taking the limit $k_j^2\to0$ eventually.
Choosing the shift such that $\kstrhat_j(z)=0$, we end up with helicity $-$ and an explicit factor $\kstr_j$ in the denominator.
We conclude that $U(k_j)$ in \Equation{Eq:limit1} gives the $-$ helicity amplitude in the limit $k_j^2\to0$, while $V(k_j)$ gives the $+$ helicity amplitude.

\section{\label{Sec:examples}Examples}
In this section, we calculate some amplitudes explicitly.
We denote them by $\Amp$ followed by an argument list consisting of numbers with super scripts indicating that the particular gluon is off-shell, or else its helicity.
So
%
\begin{equation}
\Amp(1^*,2^+,3^-,\ldots)
\end{equation}
%
is the amplitude for the process $\emptyset\to g^*gg\cdots$, where gluon $2$ has helicity~$+$, gluon $3$ has helicity~$-$, etc..
The correctness of all amplitudes presented in this section has been confirmed numerically using a computer program based on the approach presented in \cite{vanHameren:2012if} generalized to an arbitrary number of off-shell gluons.
%

\subsection{Three-point amplitudes}
The amplitudes for $\emptyset\to ggg$, which vanish for real momenta, are well-known in literature.
If all helicities are equal, they also vanish for shifted momenta, and the others are given by
%
\begin{equation}
\Amp(1^+,2^-,3^-) = \frac{\AA{23}^3}{\AA{31}\AA{12}}
\quad,\quad
\Amp(1^-,2^+,3^+) = \frac{\SSS{32}^3}{\SSS{21}\SSS{13}}
~,
\end{equation}
%
again on the condition that at least two of the momenta are shifted.
The amplitudes for $\emptyset\to g^*gg$ are calculated in \Appendix{appA} using a conventional approach, with the result that the equal-helicity amplitudes vanish, while
%
\begin{equation}
\Amp(1^*,2^+,3^-)
= \frac{1}{\kstr_1}\,\frac{\AA{31}^3}{\AA{12}\AA{23}}
= \frac{1}{\kapp_1}\,\frac{\SSS{21}^3}{\SSS{13}\SSS{32}}
~.
\end{equation}
%
Notice how both the ``mostly-plus MHV'' and ``mostly-minus MHV'' expressions can be used to represent the same amplitude, a phenomenon that also happens for the amplitudes for four on-shell gluons.

\subsubsection{$\emptyset\to gg^*g^*$}
We start with $\Amp(1^+,2^*,3^*)$.
There is only one contribution, namely
%
\begin{equation}
\Amp(1^+,2^*,3^*) = \graph{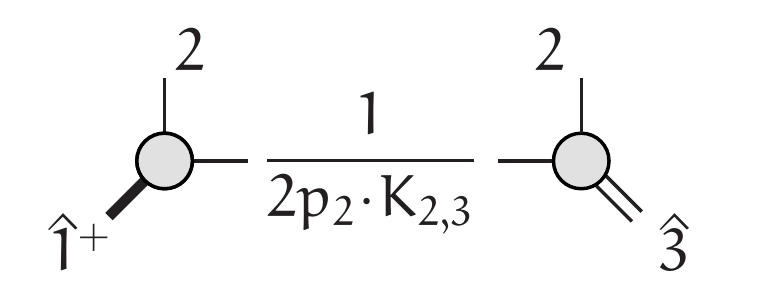}{25}{3.6}
                  = \frac{\AA{23}}{\AA{31}\AA{12}}
                    \graph{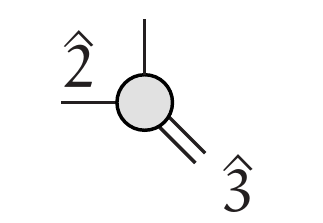}{10.5}{3}
~.
\end{equation}
%
The possible contribution of type D vanishes, because on-shell gluon $3$ will have helicity~$+$, and $\Amp(1^+,2^*,3^+)$ vanishes.
The expression above comes from \Equation{Eq:Btwo}.
%
%
So we have
%
\begin{equation}
\Amp(1^+,2^*,3^*) = \frac{\AA{23}}{\AA{31}\AA{12}}
  \,\frac{2\lid{p_2}{p_3}}{\hat{\kapp}_2\kstr_2}
= \frac{\AA{23}}{\AA{31}\AA{12}}
 \,\frac{\AA{32}\AA{23}\SSS{32}}{\AS{3|\slashk_2+\slashk_1|2}\kstr_2}
~.
\end{equation}
%
Using momentum conservation, we have $\AS{3|\slashk_2+\slashk_1|2}=-\AS{3|\slashk_3|2}=-\kstr_3\SSS{32}$, leading to
%
\begin{equation}
\Amp(1^+,2^*,3^*) = \frac{1}{\kstr_2\kstr_3}\,\frac{\AA{23}^3}{\AA{31}\AA{12}}
~.
\label{Eq:gGG}
\end{equation}
%
For the other helicity, we find
%
\begin{equation}
\Amp(1^-,2^*,3^*) = \frac{1}{\kapp_2\kapp_3}\,\frac{\SSS{32}^3}{\SSS{21}\SSS{13}}
~.
\label{Eq:gGG2}
\end{equation}
%

\subsubsection{$\emptyset\to g^*g^*g^*$}
For the process $\emptyset\to g^*g^*g^*$ we have three contributions
%
\begin{equation}
\Amp(1^*,2^*,3^*) =
  \graph{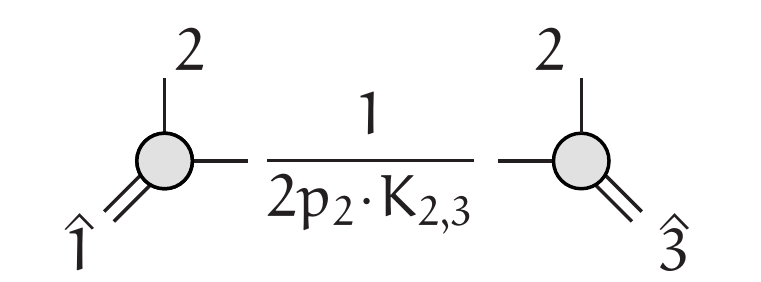}{25}{3.6}
+ \graph{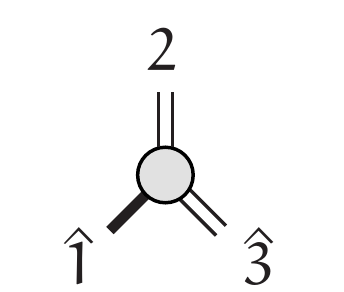}{11}{3.6}
+ \graph{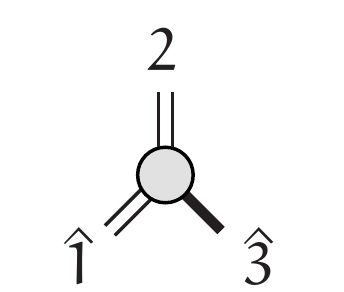}{11}{3.6}
~.
\end{equation}
%
Choosing shift vector $\frac{1}{2}\AS{1|\gamma^\mu|3}$, we can read off the first contribution from \Equation{Eq:B2off1} and \Equation{Eq:B2off2} (with the necessary permutation), so we find
%
\begin{equation}
\graph{graph20}{25}{3.6}
= \frac{\AA{12}^2\SSS{23}}{\kstr_1\AS{2|\slashk_1|3}}\,\frac{-1}{2\lid{p_2}{k_1}}\,\frac{\SSS{23}^2\AA{12}}{\kapp_3\AS{1|\slashk_3|2}}
= \frac{\AA{12}^3\SSS{32}^3}{\kapp_3\kstr_1\AS{1|\slashk_3|2}\AS{2|\slashk_1|3}\AS{2|\slashk_1|2}}
~.
\end{equation}
%
For the second contribution, we have
%
\begin{equation}
\graph{graph21}{11}{3.6}
=
\frac{1}{x_1(p_3)\kapp_1}\,\frac{1}{\kstr_2\kstrhat_3}
\,\frac{\AA{23}^3}{\AA{3\hat{k}_1}\AA{\hat{k}_12}}
~,
\end{equation}
%
where $\kstrhat_3$, $\AA{3\hat{k}_1}$ and $\AA{\hat{k}_12}$ follow from \Equation{Eq:Cbraket}.
Using also momentum conservation to write $\AS{3|\slashk_3+\slashk_1|1}=-\AS{3|\slashk_2|1}$ and $\lid{k_1}{p_3}=-\lid{(k_2+k_3)}{p_3}=-\lid{k_2}{p_3}$, we find
%
\begin{equation}
\graph{graph21}{11}{3.6}
=
\frac{\AA{23}^3\SSS{13}^3}{\kapp_1\kstr_2\AS{2|\slashk_1|3}\AS{3|\slashk_2|1}\AS{3|\slashk_2|3}}
~.
\end{equation}
%
The third term can be found similarly, with the final result
%
\begin{equation}
\Amp(1^*,2^*,3^*) = 
\frac{\AA{12}^3\SSS{32}^3}{\kapp_3\kstr_1\AS{1|\slashk_3|2}\AS{2|\slashk_1|3}\AS{2|\slashk_1|2}}
+ (231) + (312) 
~,
\end{equation}
%
where the second and third term are obtained by applying the cyclic permutations on the arguments and indices of the first term.

\subsection{Four-point amplitudes}
The amplitudes for $\emptyset\to gggg$ are of course well-known, and given by
%
\begin{equation}
\Amp(i^+,j^+,k^-,l^-) = \frac{\AA{kl}^4}{\AA{12}\AA{23}\AA{34}\AA{41}}
                      = \frac{\SSS{ij}^4}{\SSS{43}\SSS{32}\SSS{21}\SSS{14}}
~.
\end{equation}

\subsubsection{$\emptyset\to g^*ggg$}
Compact expression for $\emptyset\to g^*ggg$ were already presented in~\cite{vanHameren:2012uj}, and we will reproduce some of them here.
The amplitudes for which all on-shell gluons have the same helicity can easily be seen to vanish.
With shift vector $\frac{1}{2}\AS{1|\gamma^\mu|4}$ for example
%
\begin{equation}
\Amp(1^*,2^+,3^+,4^+)
=\graph{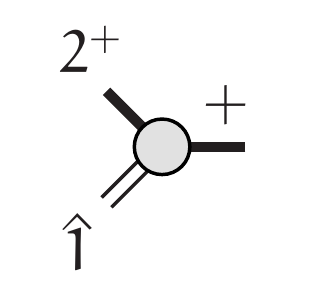}{10}{3.8}\!\!\frac{1}{K_{3,4}^2}\!\graph{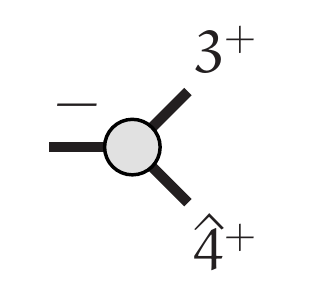}{10}{3.8}
+\graph{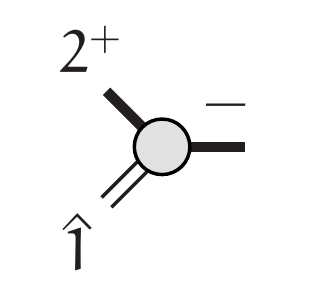}{10}{3.8}\!\!\frac{1}{K_{3,4}^2}\!\graph{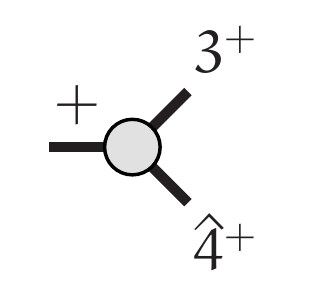}{10}{3.8}
+\graph{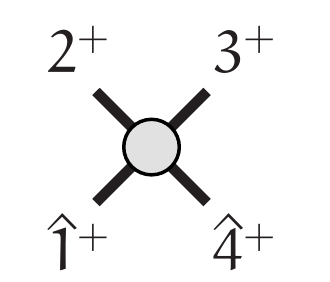}{10.5}{3.7}
,
\end{equation}
%
and for each term, there is an amplitude that vanishes.
For $\Amp(1^*,2^+,3^+,4^-)$ the only contributions that do not vanish {\it a priori\/} with shift vector $\frac{1}{2}\AS{4|\gamma^\mu|1}$ are
%
\begin{equation}
\Amp(1^*,2^+,3^+,4^-)
=\graph{graph26}{10}{3.8}\!\!\frac{1}{K_{3,4}^2}\!\graph{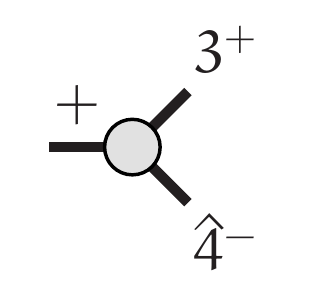}{10}{3.8}
+\graph{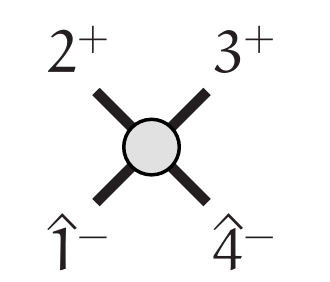}{10.5}{3.7}
.
\end{equation}
%
The first contribution, however, does vanish because of the amplitude on the right hand side, following a well known phenomenon in BCWF recursion for on-shell amplitudes.
There are two options to represent the on-shell amplitude for the second term: the ``mostly-plus MHV'' representation, and the ``mostly-minus MHV'' representation.
The latter is the hard one in this respect.
We have
%
\begin{equation}
\Amp(1^*,2^+,3^+,4^-)
=\frac{1}{x_1(p_4)\kstr_1}\,\frac{\SSS{32}^3}{\SSS{\hat{k}_1\hat{4}}\SSS{\hat{4}3}\SSS{2\hat{k}_1}}
~,
\end{equation}
%
where the shifted quantities follow from \Equation{Eq:Dbraket} with $1\leftarrow4,n\leftarrow1$.
First, we have
%
\begin{equation}
\SSS{\hat{k}_1\hat{4}}
=
\frac{\AL{4}\slashk_1(\slashp_4+\slashk_1)\AR{1}}{\sqrt{x_1}\AA{41}\AA{41}}
=\frac{2\lid{p_4}{k_1}+k_1^2}{\sqrt{x_1}\AA{41}}
=\frac{1}{\sqrt{x_1}}\,\frac{(p_2+p_3)^2}{\AA{41}}
=\frac{1}{\sqrt{x_1}}\,\frac{\AA{23}\SSS{32}}{\AA{41}}
~,
\end{equation}
%
where we used momentum conservation.
Next, also using momentum conservation, we have
\begin{equation}
\SSS{\hat{4}3} = -\SSS{3\hat{4}}
=-\frac{\SA{3|\slashp_4+\slashk_1|1}}{\AA{41}}
=\frac{\SA{3|\slashp_3+\slashp_2|1}}{\AA{41}}
=\frac{\AA{12}\SSS{23}}{\AA{41}}
~.
\end{equation}
%
And finally, again applying also momentum conservation, we find
%
\begin{equation}
\SSS{2\hat{k}_1}
= \frac{\SA{2|\slashk_1|4}}{\sqrt{x_1}\AA{14}}
=-\frac{\AS{4|\slashp_2+\slashp_3+\slashp_4|2}}{\sqrt{x_1}\AA{14}}
=-\frac{\AA{43}\SSS{32}}{\sqrt{x_1}\AA{14}}
= \frac{\AA{34}\SSS{32}}{\sqrt{x_1}\AA{41}}
~.
\end{equation}
%
Putting everything together, we get
%
\begin{equation}
\Amp(1^*,2^+,3^+,4^-)
=-\frac{1}{x_1\kstr_1}\,\frac{\SSS{32}^3\AA{41}^3x_1}{\AA{23}\SSS{32}\AA{12}\SSS{23}\AA{34}\SSS{32}}
=\frac{1}{\kstr_1}\,\frac{\AA{41}^3}{\AA{12}\AA{23}\AA{34}}
~.
\end{equation}
%
Notice that we would have found this result almost immediately had we used the ``mostly-plus MHV'' representation of the on-shell amplitude, since the only angular spinor that shifts is $\AR{\hat{k}_1}=\sqrt{x_1}\AR{1}$.

Shifting $1$ and $4$ with shift vector $\frac{1}{2}\AS{1|\gamma^\mu|4}$, there is only one contribution to $\Amp(1^*,2^+,3^-,4^+)$ that does not vanish {\it a priori\/}, namely
%
\begin{equation}
\Amp(1^*,2^+,3^-,4^+)
=
\graph{graph26}{10}{3.8}\!\!\frac{1}{K_{3,4}^2}\!\graph{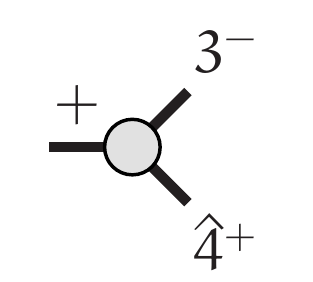}{10}{3.8}
=\frac{1}{\kstr_1}\,\frac{\AA{\hat{K}1}^3}{\AA{12}\AA{2\hat{K}}}
 \,\frac{1}{K_{3,4}^2}\,
 \frac{\SSS{\hat{K}\hat{4}}^3}{\SSS{\hat{4}3}\SSS{3\hat{K}}}
~,
\end{equation}
%
with
%
\begin{gather}
\hat{K}^\mu = p_3^\mu+p_4^\mu - \frac{z}{2}\AS{1|\gamma^\mu|4}
\quad,\quad
z = \frac{\AA{34}}{\AA{31}}
\\
\AR{\hat{4}} = \AR{4}-z\AR{1} = \frac{\AA{14}}{\AA{13}}\,\AR{3}
\quad,\quad
\AR{\hat{K}}\SL{\hat{K}}
=\AR{3}\SL{3}+\AR{\hat{4}}\SL{4}
=\AR{3}\left(\SL{3}+\frac{\AA{14}}{\AA{13}}\,\SL{4}\right)
%
~,
\end{gather}
%
leading to
%
\begin{equation}
\Amp(1^*,2^+,3^-,4^+)
=
\frac{1}{\kstr_1}\,\frac{\AA{31}^3}{\AA{12}\AA{23}}\,\frac{1}{\AA{34}\SSS{43}}
\,\frac{\SSS{34}^3}{\SSS{43}\frac{\AA{14}}{\AA{13}}\SSS{34}}
=
\frac{1}{\kstr_1}\,\frac{\AA{31}^4}{\AA{12}\AA{23}\AA{34}\AA{41}}
~.
\end{equation}
%
We see that, besides a factor $\gQCD^2/\sqrt{2}$, the amplitudes here differ a factor $-\AS{1|\slashk_1|3}/\AS{3|\slashk_1|1}\sqrt{-k_1^2}$ with the ones in~\cite{vanHameren:2012uj} for the mostly-plus cases.
For the mostly-minus cases, it turns out to be a factor $-\sqrt{-k_1^2}$.
So in both cases, it is $\sqrt{-k_1^2}$ times a phase factor.

\subsubsection{$\emptyset\to g^*g^*gg$}
Shifting $1$ and $4$ with shift vector $\frac{1}{2}\AS{1|\gamma^\mu|4}$, there is only one non-vanishing contribution to $\Amp(1^*,2^+,3^+,4^*)$, namely
%
\begin{equation}
\Amp(1^*,2^+,3^+,4^*)
=
\graph{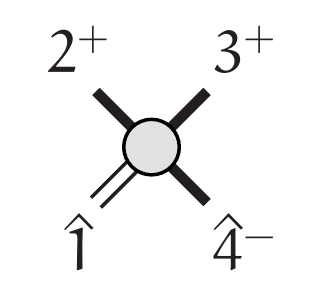}{10.5}{3.7}
=
\frac{1}{x_4(p_1)\kstr_4}\,\frac{1}{\kstr_1}
\,\frac{\AA{\hat{k}_41}^3}{\AA{12}\AA{23}\AA{3\hat{k}_4}}
=
\frac{1}{\kstr_4\kstr_1}
\,\frac{\AA{41}^3}{\AA{12}\AA{23}\AA{34}}
~.
\end{equation}
%
For $\Amp(1^*,2^+,3^-,4^*)$, there are three contributions
%
\begin{equation}
\Amp(1^*,2^+,3^-,4^*)
=
\graph{graph26}{10}{3.8}\!\!\frac{1}{K_{3,4}^2}\!\graph{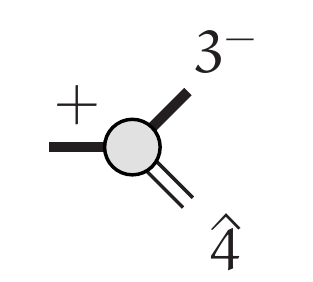}{10}{3.8}
+ \graph{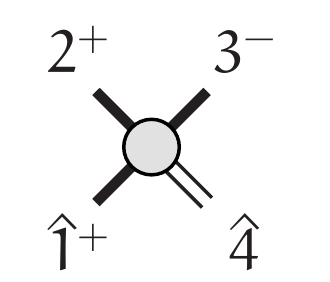}{10.5}{3.7} + \graph{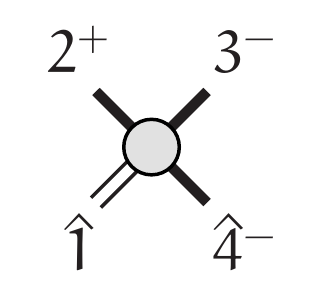}{10.5}{3.7}
~.
\end{equation}
%
The shifted momentum in the internal line of the first term is given by
%
\begin{equation}
\hat{\slashK} = \slashp_3 + \slashk_4
-z\big(\AR{1}\SL{4}+\SR{4}\AL{1}\big)
\quad,\quad
z = \frac{(p_3+k_4)^2}{\AS{1|\slashp_3+\slashk_4|4}}
~.
\end{equation}
%
Other quantities do not shift (\Equation{Eq:shifts}), so
%
\begin{multline}
\graph{graph26}{10}{3.8}\!\!\frac{1}{K_{3,4}^2}\!\graph{graph34}{10}{3.8}
=
\frac{1}{\kstr_1}\,\frac{\AA{\hat{K}1}^3}{\AA{12}\AA{2\hat{K}}}
\,\frac{1}{(p_3+k_4)^2}\,\frac{1}{\kapp_4}\,\frac{\SSS{\hat{K}4}^3}{\SSS{43}\SSS{3\hat{K}}}
\\
=
\frac{1}{\kstr_1\kapp_4}\,\frac{\AS{1|\hat{\slashK}|4}^3}{\AS{2|\hat{\slashK}|3}\AA{12}\SSS{43}(p_3+k_4)^2}
=\frac{1}{\kstr_1\kapp_4}
\,\frac{-\AS{1|\slashp_3+\slashk_4|4}^4}
       {\AS{2|\slashk_1|4}\AS{1|\slashk_4|3}\AA{12}\SSS{43}(p_3+k_4)^2}
~,
\end{multline}
%
where we used \Equation{Eq:AppB1} with $K^\mu=p_3^\mu+k_4^\mu$, and we used momentum conservation.
For the second term we have
%
\begin{equation}
\graph{graph35}{10.5}{3.7}
=
\frac{1}{x_1(p_4)\kapp_1}\,\frac{1}{\kstrhat_4}\,\frac{\AA{34}^3}{\AA{\hat{k}_12}\AA{23}\AA{4\hat{k}_1}}
=
\frac{1}{\kapp_1}\,\frac{\AA{34}^3\SSS{14}^3}{\AS{4|\slashk_4+\slashk_1|1}\AS{2|\slashk_1|4}\AS{4|\slashk_1|4}\AA{23}}
~,
\end{equation}
%
where we took the shifted quantities from \Equation{Eq:Cbraket}.
The third term is
%
\begin{equation}
\graph{graph36}{10.5}{3.7}
=
\frac{1}{x_n(p_1)\kstr_n}\,\frac{1}{\kapphat_1}
\,\frac{\SSS{21}^3}{\SSS{1\hat{k}_4}\SSS{\hat{k}_43}\SSS{32}}
=
\frac{1}{\kstr_4}\,\frac{\SSS{21}^3\AA{14}^3}{\AS{4|\slashk_4+\slashk_1|1}\AS{1|\slashk_4|3}\AS{1|\slashk_4|1}\SSS{32}}
~.
\end{equation}
%
The second and third term can be combined as follows
%
\begin{equation}
\graph{graph35}{10.5}{3.7} + \graph{graph36}{10.5}{3.7}
= \frac{1}{\AA{41}\kapp_1+\SSS{41}\kstr_4}
\left(\frac{\SSS{41}}{\kapp_1}\,F(k_1,k_4) + \frac{\AA{41}}{\kstr_4}\,F_*(k_1,k_4)\right)
~,
\end{equation}
%
with
%
\begin{equation}
F(p_1,p_4)
=F_*(p_1,p_4)
=\Amp(1^+,2^+,3^-,4^-)
~,
\end{equation}
%
$F(p_1,p_2)$ giving the ``mostly-plus MHV'' representation, and $F_*(p_1,p_4)$ giving the ``mostly-minus MHV'' representation.
The first term, multiplied with $\kstr_1\kapp_4$, gives $\Amp(1^-,2^+,3^-,4^+)$ when $k_1\to p_1$ and $k_4\to p_4$, in the ``mostly-plus MHV'' representation.

There is only one non-vanishing contribution to $\Amp(1^*,2^+,3^*,4^+)$, namely
%
\begin{equation}
\Amp(1^*,2^+,3^*,4^+) = \graph{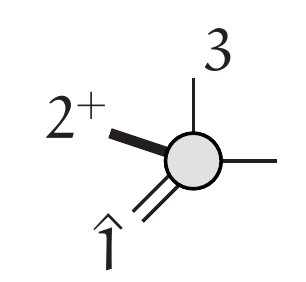}{9.5}{3.6}
                    \frac{1}{2\lid{p_3}{K_{3,4}}}
                    \graph{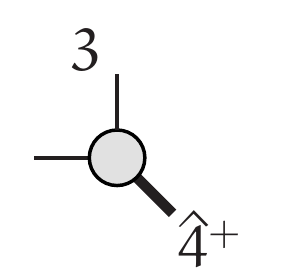}{9.3}{3.6}
= \frac{1}{\kstr_1\kstr_3}\,\frac{\AA{13}^4}{\AA{12}\AA{23}\AA{34}\AA{41}}
~,
\end{equation}
%
which follows directly from \Equation{Eq:Btwo} and \Equation{Eq:gGG}.

Again using shift vector $\frac{1}{2}\AS{1|\gamma^\mu|4}$, there are three contributions to $\Amp(1^*,2^-,3^*,4^+)$ that do not vanish {\it a priori\/}
%
\begin{equation}
\Amp(1^*,2^-,3^*,4^+)
=
\!\!\graph{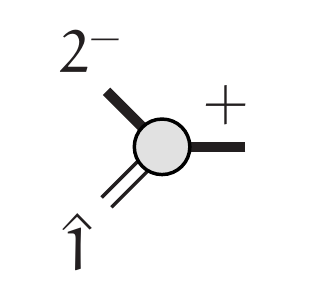}{10}{3.8}\!\!\frac{1}{K_{3,4}^2}\!\graph{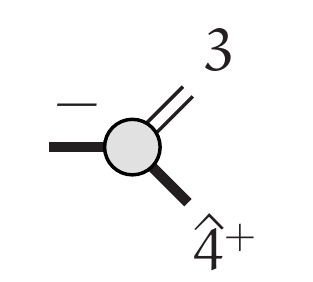}{10}{3.8}
\!\!\!+\graph{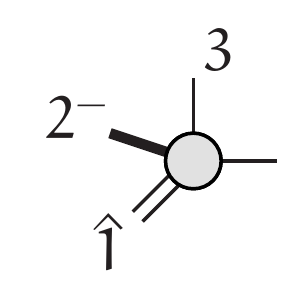}{9.5}{3.6}\frac{1}{2\lid{p_3}{K_{3,4}}}\graph{graph37}{9.3}{3.6}
\!\!\!+\graph{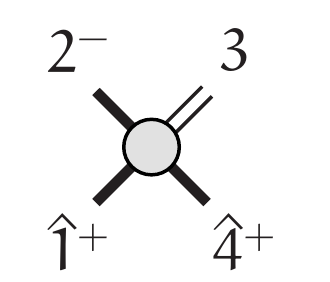}{10.5}{3.7}
\!\!\!.
\end{equation}
%
The shifted momentum in the internal line of the first term is given by
%
\begin{equation}
\hat{\slashK} = \slashk_3 + \slashp_4
-z\big(\AR{1}\SL{4}+\SR{4}\AL{1}\big)
\quad,\quad
z = \frac{(k_3+p_4)^2}{\AS{1|\slashk_3|4}}
~.
\end{equation}
%
Other quantities do not shift (\Equation{Eq:shifts}), so
%
\begin{multline}
\graph{graph39}{10}{3.8}\!\!\frac{1}{K_{3,4}^2}\!\graph{graph40}{10}{3.8}
=
\frac{1}{\kstr_1}\,\frac{\AA{12}^3}{\AA{2\hat{K}}\AA{\hat{K}1}}
\,\frac{1}{(k_3+p_4)^2}
\,\frac{1}{\kapp_3}\,\frac{\SSS{43}^3}{\SSS{3\hat{K}}\SSS{\hat{K}4}}
\\
=\frac{1}{\kstr_1\kapp_3}\,\frac{\AA{12}^3\SSS{43}^3}
         {\AS{2|\hat{\slashK}|4}\AS{1|\hat{\slashK}|3}(k_3+p_4)^2}
=\frac{1}{\kstr_1\kapp_3}\,\frac{\AA{12}^3\SSS{43}^3}
         {\AS{2|\slashk_3|4}\AS{1|\slashk_3+\slashp_4|3}(k_3+p_4)^2}
~.
\end{multline}
%
For the second term we find, using \Equation{Eq:Btwo}, \Equation{Eq:Btwo2} and  \Equation{Eq:gGG2},
%
\begin{equation}
\graph{graph41}{9.5}{3.6}\frac{1}{2\lid{p_3}{K_{3,4}}}\graph{graph37}{9.3}{3.6}
=
\frac{\AA{13}^3\SSS{13}^3}{\AA{34}\AA{41}\AS{1|\slashk_3+\slashp_4|3}\AS{3|\slashk_1+\slashp_4|1}\SSS{32}\SSS{21}}
~.
\end{equation}
%
This term vanishes when either gluon $1$ or gluon $3$ becomes on-shell.
The third term gives
%
\begin{equation}
\graph{graph42}{10.5}{3.7}
= \frac{1}{x_1(p_4)\kapp_1}\,\frac{1}{\kstr_3}\,
\frac{\AA{23}^3}{\AA{\hat{k}_12}\AA{3\hat{4}}\AA{\hat{4}\hat{k}_1}}
=\frac{1}{\kapp_1\kstr_3}\,\frac{\AA{23}^3\SSS{14}^3}
         {\AS{2|\slashk_1|4}\AS{3|\slashk_1+\slashp_4|1}(k_1+p_4)^2}
~,
\end{equation}
%
where we took the shifted quantities from \Equation{Eq:Cbraket}.

\subsubsection{$\emptyset\to g^*g^*g^*g$}
Using shift vector $\frac{1}{2}\AS{1|\gamma^\mu|4}$, there are four non-vanishing contributions to $\Amp(1^*,2^*,3^+,4^*)$.
The first is
%
\begin{multline}
\graph{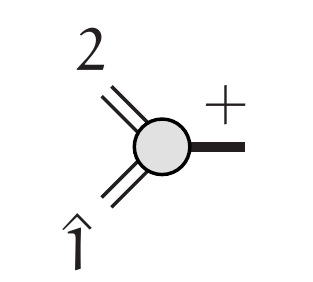}{10}{3.8}\!\!\frac{1}{K_{3,4}^2}\!\graph{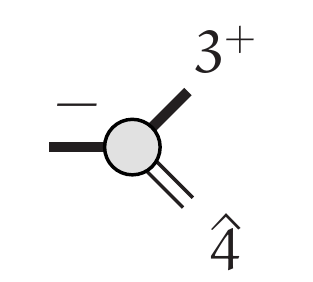}{10}{3.8}
=
\frac{1}{\kstr_1\kstr_2}\,\frac{\AA{12}^3}{\AA{2\hat{K}}\AA{\hat{K}1}}
\,\frac{1}{(p_3+k_4)^2}\,\frac{1}{\kapp_4}\,\frac{\SSS{43}^3}{\SSS{3\hat{K}}\SSS{\hat{K}4}}
\\
=\frac{1}{\kstr_1\kstr_2\kapp_4}\,\frac{\AA{12}^3\SSS{43}^3}
         {\AS{1|\slashK|3}\AS{2|\slashK|4}(p_3+k_4)^2}
=\frac{1}{\kstr_1\kstr_2\kapp_4}\,\frac{\AA{12}^3\SSS{43}^3}
         {\AS{1|\slashk_4|3}\AS{2|\slashp_3+\slashk_4|4}(p_3+k_4)^2}
~.
\label{Eq:GGGg1}
\end{multline}
%
The second is
%
\begin{multline}
\graph{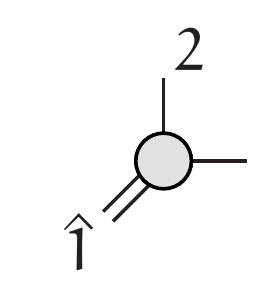}{9.0}{3.6}\frac{1}{2\lid{p_2}{K_{3,4}}}\graph{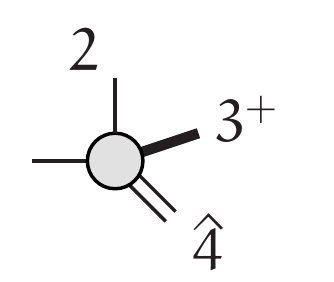}{10.8}{3.6}
=\frac{\AA{12}^2\SSS{24}}{\kstr_1\AS{2|\slashk_1|4}}\,\frac{1}{2\lid{p_2}{k_1}}
\,\frac{1}{\kstrhat_2\kstrhat_4}\,\frac{\AA{42}^3}{\AA{23}\AA{34}}
\\
=\frac{1}{\kstr_1}\,\frac{\AA{12}^3\AA{24}^3\SSS{42}^3}
 {\AS{2|\slashk_1|4}\AS{2|\slashk_1|2}\AS{2|\slashk_2+\slashk_1|4}
  \big(\AS{4|\slashk_4|2}\AA{12}+\AS{2|\slashk_1|2}\AA{14}\big)\AA{23}\AA{34}}
~.
\label{Eq:GGGg2}
\end{multline}
%
Next, we have
%
\begin{multline}
\graph{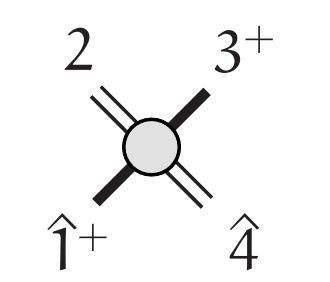}{10.5}{3.7}
= \frac{1}{x_1(p_4)\kapp_1}\,\frac{1}{\kstr_2\kstrhat_4}
 \,\frac{\AA{24}^4}{\AA{\hat{k}_12}\AA{23}\AA{34}\AA{4\hat{k}_1}}
\\
= \frac{1}{\kapp_1\kstr_2}\,\frac{\AA{24}^4\SSS{14}^3}
  {\AS{4|\slashk_1+\slashk_4|1}\AS{2|\slashk_1|4}\AS{4|\slashk_1|4}\AA{23}\AA{34}}
~.
\label{Eq:GGGg3}
\end{multline}
%
And finally
%
\begin{equation}
\graph{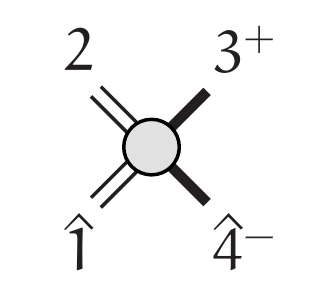}{10.5}{3.7}
= \Amp_1 + \Amp_2 + \Amp_3
~,
\end{equation}
%
with
%
\begin{multline}
\Amp_1 = \frac{1}{x_4(p_1)\kstr_4}
\,\frac{1}{\kstr_2\kapphat_1}
\,\frac{-\AS{2|\hat{\slashk}_4+\hat{\slashk}_1|1}^4}
       {\AS{3|\slashk_2|1}\AS{2|\hat{\slashk}_1|\hat{k}_4}\AA{23}\SSS{1\hat{k}_4}(\hat{k}_4+\hat{k}_1)^2}
\\
= \frac{1}{\kstr_2\kstr_4}
\,\frac{-\AS{2|\slashk_1+\slashk_4|1}^4\AA{41}^3}
  {\AS{4|\slashk_1+\slashk_4|1}\AS{3|\slashk_2|1}\AA{2|(\slashk_1+\slashk_4)\slashk_4|1}\AA{23}\AS{1|\slashk_4|1}(k_1+k_4)^2}
\\
= \frac{1}{\kstr_2\kstr_4}
\,\frac{\AS{2|\slashk_2+\slashp_3|1}^4\AA{41}^3}
  {\AS{4|\slashk_1+\slashk_4|1}\AS{3|\slashk_2|1}\AA{2|(\slashk_2+\slashp_3)\slashk_4|1}\AA{23}\AS{1|\slashk_4|1}(k_2+p_3)^2}
~,
\label{Eq:GGGg4}
\end{multline}
%
\begin{multline}
\Amp_2 = \frac{1}{x_4(p_1)\kstr_4}\,
\frac{1}{\kapp_2}\,\frac{\AA{\hat{k}_41}^3\SSS{21}^3}{\AS{1|\hat{\slashk}_1+\slashk_2|2}\AS{3|\slashk_2|1}\AS{1|\slashk_2|1}\AA{3\hat{k}_4}}
\\
=\frac{1}{\kapp_2\kstr_4}\,\frac{\AA{41}^3\SSS{21}^3}{\AS{1|\slashk_1+\slashk_2|2}\AS{3|\slashk_2|1}\AS{1|\slashk_2|1}\AA{34}}
~,
\label{Eq:GGGg5}
\end{multline}
%
\begin{multline}
\Amp_3 =
\frac{1}{x_4(p_1)\kstr_4}
\,\frac{1}{\kstr_1}\,\frac{\SSS{32}^3\AA{21}^3}{\AS{1|\hat{\slashk}_1+\slashk_2|2}\AS{2|\hat{\slashk}_1|\hat{k}_4}\AS{2|\hat{\slashk}_1|2}\SSS{\hat{k}_43}}
\\
=\frac{1}{\kstr_1\kstr_4}\,\frac{\SSS{32}^3\AA{12}^3\AA{41}^3}
  {\AS{1|\slashk_1+\slashk_2|2}\AA{2|(\slashk_1+\slashk_4)\slashk_4|1}
   \big(\AS{2|\slashk_1|2}\AA{41}+\AS{4|\slashk_4|2}\AA{21}\big)\AS{1|\slashk_4|3}}
~.
\label{Eq:GGGg6}
\end{multline}
%
We see that (\ref{Eq:GGGg1}) gives $\Amp(1^-,2^-,3^+,4^+)$ when all momenta go on-shell, while (\ref{Eq:GGGg3}) and (\ref{Eq:GGGg4}) combine to give $\Amp(1^+,2^-,3^+,4^-)$, and (\ref{Eq:GGGg5}) and (\ref{Eq:GGGg6}) combine to give $\Amp(1^-,2^+,3^+,4^-)$.

\section{\label{Sec:MHV}MHV-amplitudes}
It is well-known that on-shell amplitudes for which all gluons, or all but one gluon, have the same helicity vanish, and that on-shell amplitudes for which all but two gluons have the same helicity are given by the simple expressions
%
\begin{align}
\Amp(i^-,j^-,(\textrm{the rest})^+)
&=\frac{\AA{p_ip_j}^4}{\AA{p_1p_2}\AA{p_2p_3}\cdots\AA{p_{n-2}p_{n-1}}\AA{p_{n-1}p_n}\AA{p_np_1}}
\\
\Amp(i^+,j^+,(\textrm{the rest})^-)
&=\frac{\SSS{p_jp_i}^4}{\SSS{p_1p_n}\SSS{p_np_{n-1}}\SSS{p_{n-1}p_{n-2}}\cdots\SSS{p_3p_2}\SSS{p_2p_1}}
~.
\end{align}
%
These are the so-called maximum-helicity-violating (MHV) amplitudes.
A similar phenomenon appears for amplitudes with off-shell gluons.

It is rather easy to see that amplitudes for one off-shell gluon and for which all on-shell gluons have the same helicity vanish
%
\begin{equation}
\Amp(1^*,2^+,3^+,\ldots,n^+) = \Amp(1^*,2^-,3^-,\ldots,n^-) = 0
~.
\end{equation}
%
We already saw for that four-point amplitudes with one off-shell gluon, say gluon $1$, are given by the MHV formula, augmented with a factor $1/\kapp_1$ if two on-shell gluons have helicity $-$, and a factor $1/\kstr_1$ if two on-shell gluons have helicity $+$.
Let us now consider general $n$-point amplitudes.
If one of the on-shell gluons has helicity $-$ while all other on-shell gluons have helicity $+$, and the $-$ helicity gluon is adjacent to the off-shell gluon, the amplitude is, using shift vector $\frac{1}{2}\AS{n|\gamma^\mu|1}$, given by two contributions that do not vanish {\it a priori\/}
%
\begin{equation}
\Amp(1^*,2^+,\ldots,(n-1)^+,n^-)
=\graph{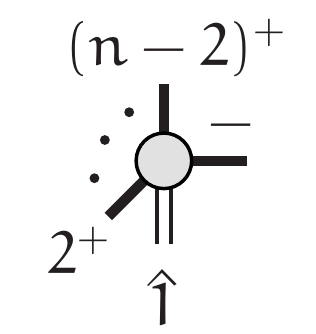}{11.0}{5.0}\!\!\frac{1}{(p_{n-1}+p_n)^2}\!\graph{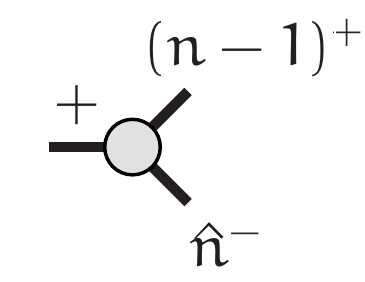}{12.0}{3.6}
+\graph{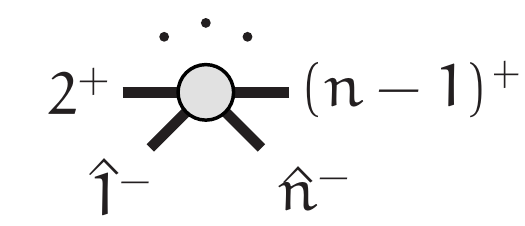}{17.5}{3.0}
~.
\end{equation}
%
The first contribution, however, also vanishes, while the second gives
%
\begin{align}
\Amp(1^*,2^+,\ldots,(n-1)^+,n^-)
&=\frac{1}{x_1(p_n)\kstr_1}\,\frac{\AA{p_n\hat{k}_1}^4}{\AA{\hat{k}_1p_2}\AA{p_2p_3}\cdots\AA{p_{n-1}p_n}\AA{p_n\hat{k}_1}}
\notag\\
&=\frac{1}{\kstr_1}\,\frac{\AA{p_np_1}^4}{\AA{p_1p_2}\AA{p_2p_3}\cdots\AA{p_{n-1}p_n}\AA{p_np_1}}
~,
\label{Eq:MHV1}
\end{align}
%
where the shifted momentum is given by the relevant permutation of \Equation{Eq:Dbraket}.
So we find the MHV formula.
If there is one gluon with helicity $+$ between the off-shell gluon and the gluon with helicity $-$, we use shift vector $\frac{1}{2}\AS{1|\gamma^\mu|n}$, and the only non-vanishing contribution is given by
%
\begin{multline}
\Amp(1^*,2^+,\ldots,(n-2)^+,(n-1)^-,n^+)
=\graph{graph51}{11.0}{5.0}\!\!\frac{1}{(p_{n-1}+p_n)^2}\!\graph{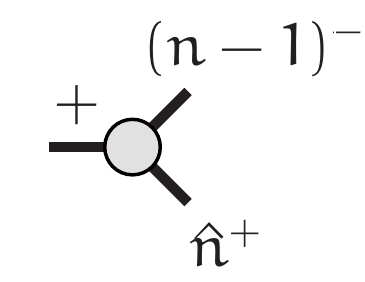}{12.0}{3.6}
\\
=\frac{1}{\kstr_1}\,\frac{\AA{\hat{K}p_1}^4}{\AA{p_1p_2}\AA{p_2p_3}\cdots\AA{p_{n-2}\hat{K}}\AA{\hat{K}p_1}}\,
\frac{1}{\AA{p_{n-1}p_n}\SSS{p_np_{n-1}}}
\,\frac{\SSS{\hat{K}\hat{p}_n}^3}{\SSS{\hat{p}_np_{n-1}}\SSS{p_{n-1}\hat{K}}}
~,
\end{multline}
%
with shifted quantities
%
\begin{equation}
\SR{\hat{p}_n} = \SR{p_n}
\quad,\quad
\SR{\hat{K}} = \SR{p_{n-1}}+\frac{\AA{p_1p_n}}{\AA{p_1p_{n-1}}}\,\SR{p_n}
\quad,\quad
\AR{\hat{K}} = \AR{p_{n-1}}
~,
\end{equation}
%
and where we used \Equation{Eq:MHV1} for the amplitude on the left side.
We find the MHV formula with $\AA{p_1p_{n-1}}^4$ in the numerator.
If there are more than one $+$ helicity gluons between the off-shell gluon and the $-$ helicity gluon, only the BCFW graph with $\Amp(\hat{K}^-,(n-1)^+,\hat{n}^+)$ on the r.h.s.\ contributes, and we again find the MHV formula.
In fact, this mechanism is completely analogous to the proof using BCFW recursion of the MHV formulas for fully on-shell amplitudes.
The same works for the amplitudes with only on gluon with helicity $+$ while the others have helicity $-$.
So we find eventually
%
\begin{align}
\Amp(1^*,i^-,(\textrm{the rest})^+)
&=\frac{1}{\kstr_1}
\,\frac{\AA{p_1p_i}^4}{\AA{p_1p_2}\AA{p_2p_3}\cdots\AA{p_{n-2}p_{n-1}}\AA{p_{n-1}p_n}\AA{p_np_1}}
\\
\Amp(1^*,i^+,(\textrm{the rest})^-)
&=\frac{1}{\kapp_1}
\,\frac{\SSS{p_ip_1}^4}{\SSS{p_1p_n}\SSS{p_np_{n-1}}\SSS{p_{n-1}p_{n-2}}\cdots\SSS{p_3p_2}\SSS{p_2p_1}}
~.
\end{align}
%
In a similar manner, it is found that the amplitudes for two off-shell gluons with all on-shell gluons having the same helicity are also given by MHV formulas.
Only the contribution in the case with one on-shell gluon between two off-shell gluons is different.
It is given by
%
\begin{equation}
\Amp(1^*,2^+,\ldots,(n-2)^+,(n-1)^*,n^+)
=\graph{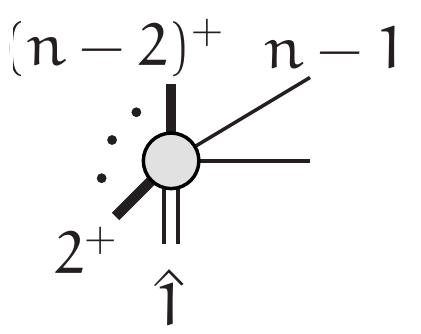}{14.0}{5.0}\!\!\frac{1}{2\lid{p_{n-1}}{p_n}}\!
\graph{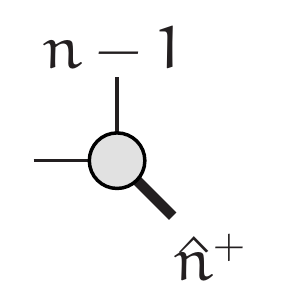}{10.0}{3.6}
~,
\end{equation}
%
leading to the MHV formula.
For the general case, we find
%
\begin{align}
\Amp(1^*,i^*,(\textrm{the rest})^+)
&=\frac{1}{\kstr_1\kstr_i}
\,\frac{\AA{p_1p_i}^4}{\AA{p_1p_2}\AA{p_2p_3}\cdots\AA{p_{n-2}p_{n-1}}\AA{p_{n-1}p_n}\AA{p_np_1}}
\\
\Amp(1^*,i^*,(\textrm{the rest})^-)
&=\frac{1}{\kapp_1\kapp_i}
\,\frac{\SSS{p_ip_1}^4}{\SSS{p_1p_n}\SSS{p_np_{n-1}}\SSS{p_{n-1}p_{n-2}}\cdots\SSS{p_3p_2}\SSS{p_2p_1}}
~.
\end{align}
%

\section{\label{Sec:summary}Summary}
We presented BCFW recursion for tree-level multi-gluon amplitudes with an arbitrary number of off-shell external gluons.
We calculated several helicity amplitudes for up to four external gluons with up to three of them off-shell.
Also, we derived formulas for maximum-helicity-violating amplitudes for an arbitrary number of external gluons with up to two of them off-shell.

\subsection*{Acknowledgments}
The author would like to thank P.~Kotko and K.~Kutak for useful discussions and comments.
All graphs were drawn with {\sc JaxoDraw}.
This work is (partly) supported by Polish National Science Centre Grant No.\ DEC-2011/03/B/ST2/02632.

\providecommand{\href}[2]{#2}\begingroup\raggedright\endgroup

\begin{appendix}
\section{\label{appA}The amplitudes for $\emptyset\to g^*gg$}
In order to calculate the amplitude for the process $\emptyset\to g^*gg$ we choose the polarization vectors of the on-shell gluons such that their inner product with $p_1^\mu$ vanishes, so
%
\begin{equation}
\vep_{i+}^\mu = \frac{\AS{1|\gamma^\mu|i}}{\sqrt{2}\AA{1i}}
\quad,\quad
\vep_{i-}^\mu = \frac{\AS{i|\gamma^\mu|1}}{\sqrt{2}\SSS{i1}}
~,
\end{equation}
%
for $i=2,3$.
Consequently, only the Feynman graph with the triple gluon vertex contributes, and with only one term, so we have
%
\begin{equation}
\Amp(1^*,2,3) = \frac{1}{\kapp_1\kstr_1}\,\frac{1}{\sqrt{2}}\,
                \lid{(p_2-p_3)}{(\sqrt{2}\,p_1)}\,\lid{\vep_2}{\vep_3}
~.
\end{equation}
%
Now we also immediately see that
%
\begin{equation}
\Amp(1^*,2^+,3^+) = \Amp(1^*,2^-,3^-) = 0
~,
\end{equation}
%
while, using momentum conservation $\lid{(p_2-p_3)}{p_1}=\lid{(2p_2+k_1)}{p_1}=2\lid{p_2}{p_1}$, 
%
\begin{equation}
\Amp(1^*,2^+,3^-) =
\frac{2\lid{p_2}{p_1}}{\kapp_1\kstr_1}\,\lid{\vep_{2+}}{\vep_{3-}}
=
\frac{2\lid{p_2}{p_1}\SSS{12}\AA{13}}{\kapp_1\kstr_1\AA{12}\SSS{31}}
=
-\frac{\AA{13}\SSS{32}}{\kstr_1\SSS{12}}\,\frac{\SSS{21}^3}{\kapp_1\SSS{13}\SSS{32}}
~.
\end{equation}
%
Using momentum conservation again, we have
%
\begin{equation}
\AA{13}\SSS{32}
=-\AS{1|\slashk_1+\slashp_2|2}
=-\AS{1|\slashk_1|2}
=-\SSS{12}\,\kstr_1
~,
\end{equation}
%
so
%
\begin{equation}
\Amp(1^*,2^+,3^-) = \frac{1}{\kapp_1}\,\frac{\SSS{21}^3}{\SSS{13}\SSS{32}}
                  = \frac{1}{\kstr_1}\,\frac{\AA{31}^3}{\AA{12}\AA{23}}
~.
\end{equation}
%
The second form can easily be checked to be correct too.

\section{\label{App:relations}Useful relations}

\subsection{\label{App:direction}}
For any momentum $k^\mu$, a {\em direction\/} $p^\mu$ satisfying $p^2=0$ and $\lid{p}{k}=0$ can be constructed as follows:
%
\begin{equation}
k^\mu = \Big(\,E\,,\,\vec{k}\,\Big)
\quad\Rightarrow\quad
p^\mu = \Bigg(\big|\vec{k}\big|\,,\,\frac{E}{\big|\vec{k}\big|}\,\vec{k}+\vec{q}\Bigg)
~,
\end{equation}
%
where $\vec{q}$ satisfies the relations
%
\begin{equation}
\vec{k}\!\cdot\!\vec{q} = 0
\quad,\quad
\vec{q}\!\cdot\!\vec{q} = -k^2
~.
\end{equation}
%
The latter requires $\vec{q}$ to be imaginary if $k^\mu$ is time-like.

\subsection{\label{App:spinors}Spinors}
The spinors employed in this work may conveniently be considered four-component vectors with two vanishing components:
%
\begin{align}
\SR{p} &= \begin{pmatrix}L(p)\\\mathbf{0}\end{pmatrix}
\quad,\quad
L(p) = \frac{1}{\sqrt{|p_0+p_3|}}\begin{pmatrix}-p_1+\imag p_2\\ p_0+p_3\end{pmatrix}
\\
\AR{p} &= \begin{pmatrix}\mathbf{0}\\ R(p)\end{pmatrix}
\quad,\quad
R(p) = \frac{\sqrt{|p_0+p_3|}}{p_0+p_3}\begin{pmatrix}p_0+p_3\\p_1+\imag p_2\end{pmatrix}
~.
\end{align}
%
The ``dual'' spinors are defined as
%
\begin{equation}
\SL{p} = \big(\,(\EuScript{E}L(p))^T\,,\,\mathbf{0}\,\big)
\quad,\quad
\AL{p} = \big(\,\mathbf{0}\,,\,(\EuScript{E}^TR(p))^T\,\big)
\quad,\quad\textrm{where}\quad
\EuScript{E} = \begin{pmatrix}0&1\\-1&0\end{pmatrix}
~.
\end{equation}
%
Notice that the definition of the ``dual'' spinors does not involve complex conjugation and that all spinors are well defined for complex momenta.
Defined as such, their dyadic products satisfy the relation
%
\begin{equation}
\AR{p}\SL{p} + \SR{p}\AL{p} = \slashp = \gamma_\mu p^\mu
~
\end{equation}
%
where the $\gamma$-matrices are in the Weyl representation with
%
\begin{equation}
\gamma^5 \equiv \imag\gamma^0\gamma^1\gamma^2\gamma^3
=\begin{pmatrix}-\mathbf{1}&\mathbf{0}\\\mathbf{0}&\mathbf{1}\end{pmatrix}
~.
\end{equation}
%
The following relations and notation are intensively used in this paper.
Here, we will use the convention that $p^\mu,q^\mu,r^\mu$ are light-like, while $k^\mu$ is not necessarily light-like.
%
\begin{gather}
\AL{p}\SR{q} = \SL{p}\AR{q} = 0
\\
\AL{p}\AR{p} = \SL{p}\SR{p} = 0
\\
\slashp\AR{p} = \slashp\SR{p}=0 \quad,\quad \AL{p}\slashp = \SL{p}\slashp = 0
\\
\AA{pq}\equiv\AL{p}\AR{q} \quad,\quad \SSS{pq}\equiv\SL{p}\SR{q}
\\
\AA{qp}=-\AA{pq} \quad,\quad \SSS{qp}=-\SSS{pq}
\\
\AA{pq}\SSS{qp} = 2\lid{p}{q}
\\
\AS{p|\slashk|q} = \SA{q|\slashk|q}
\\
\AS{p|\slashr|q} = \AA{pr}\SSS{rq} 
\end{gather}

\subsection{\label{App:Schouten}Schouten identity}
The Schouten identity can conveniently be expressed as a completeness relation.
For any $p^\mu,r^\mu$ with $p^2=r^2=0$ and $\lid{p}{r}\neq0$ we have
%
\begin{equation}
  \frac{\AR{r}\AL{p}}{\AA{pr}} + \frac{\AR{p}\AL{r}}{\AA{rp}}
 +\frac{\SR{r}\SL{p}}{\SSS{pr}} + \frac{\SR{p}\SL{r}}{\SSS{rp}}
=1
~.
\label{Eq:Schouten}
\end{equation}
%
A simple application is the following.
Inserting $1$ strategically, we find for $\kappa$ from \Equation{Eq:defkappas}
%
\begin{equation}
\kappa = \frac{\AS{q|\slashk|p}}{\AA{qp}}
 = \frac{\AS{q|1\slashk|p}}{\AA{qp}}
       = \frac{\AA{qr}\AS{p|\slashk|p}}{\AA{pr}\AA{qp}}
       + \frac{\AA{qp}\AS{r|\slashk|p}}{\AA{rp}\AA{qp}}
       = \frac{\AS{r|\slashk|p}}{\AA{rp}}
~,
\end{equation}
%
where we used the fact that $\AS{p|\slashk|p}=2\lid{p}{k}=0$.
The same can be shown for $\kstr$.

\subsection{}
The following relation holds for any $K^\mu$.
Introducing the vector $e^\mu = \frac{1}{2}\AS{1|\gamma^\mu|2}$ we have
%
\begin{equation}
\AS{3|\slashK|2}\AS{1|\slashK|4}
=\AS{3|\slashK\slashe\slashK|4}
=2\lid{K}{e}\AS{3|\slashK|4} - K^2\AS{3|\slashe|4}
=\AS{1|\slashK|2}\AS{3|\slashK|4} - K^2\AA{31}\SSS{24}
\label{Eq:AppB1}
~.
\end{equation}

\end{appendix}

\end{document}